\documentstyle[aps,eqsecnum,pre,amssymb,graphicx]{revtex}
 \input epsf  
\begin{document} 
\title{\bf Effects of weak surface fields on the  density profiles and   adsorption of a confined fluid near bulk criticality} 
\author{  A. Macio\l ek,$^1$  R. Evans, $^{2,3}$    and  N. B. Wilding $^4$\\    
 {\small $^1$\it  Institute of Physical Chemistry, Polish Academy of Sciences,}\\  
 {\small \it      Department III, Kasprzaka 44/52, PL-01-224 Warsaw, Poland}\\ 
 {\small $^2$ \it H.H.Wills Physics Laboratory, University of  Bristol,   }\\
 {\small \it Bristol BS8 1TL, UK } \\
 {\small $^3$  \it Max-Planck-Institut f\"{u}r Metallforschung,}\\
 {\small \it Heisenbergstr. 3, D-70569 Stuttgart, Germany,}\\ 
 {\small $^4$  \it Department  of Physics, University of Bath}\\
 {\small \it Bath BA27AY,U.K.}} 
\maketitle  
\begin{abstract}   
The density profile $\rho(z)$ and Gibbs adsorption $\Gamma$ of a near-critical 
fluid confined between two identical planar walls is studied by means of
 Monte Carlo simulation  and by density functional theory 
for a Lennard-Jones fluid.
By reducing the  strength of wall-fluid interactions relative to fluid-fluid
interactions we observe
a crossover from behaviour characteristic of the normal
 surface universality class, strong critical adsorption,
 to   behaviour characteristic of
a 'neutral' wall. The crossover  is  reminiscent of that which occurs
 near the ordinary  surface transition in Ising films subject to vanishing
surface fields. For the  'neutral' wall
 $\rho(z)$, away from the walls, is almost
 constant throughout the slit capillary and  gives rise to a $\Gamma$
 that is constant along the critical isochore. The same 'neutral' wall yields
a line of capillary coexistence that is almost
 identical to the bulk coexistence line. In the crossover regime we observe
features in $\rho(z)$ similar to those found  
in the magnetisation profile of the critical Ising film subject
 to weak surface fields, namely two smooth maxima, located away from the walls,
which merge into a single maximum at midpoint as the strength of the wall-fluid
interaction is reduced or as the distance between walls is decreased.
We discuss similarities and differences between the surface critical 
behaviour of fluids and of Ising magnets.
\end{abstract}  
\draft  
\section{Introduction} 
\label{sec:intr} 
When a fluid, a fluid mixture, a binary alloy or an Ising magnet that is close    
to a critical point in  bulk  is brought to a surface its properties are
 altered significantly. In particular the  bulk universality class is divided into
 several surface universality classes which depend on whether  the tendency  to order
at the surface is increased or decreased w.r.t that in bulk~\cite{binder:83:0,diehl:86:0}. 
For the case of Ising magnets the phenomenology is well-understood: the critical
behavior of the semi-infinite system is governed  by a surface scaling
field  $c$ which describes the
enhancement of interactions in the surface layer. $c>0$ corresponds to a reduced
tendency  to order in the surface  and $c=\infty$ defines  the fixed point
of the renormalization group transformation of the ordinary transition.
Similarly  $c=-\infty$ defines the fixed point 
for the extraordinary transition and $c=c^*$ corresponds 
to the special transition. Such a classification pertains to vanishing 
(external) surface field $h_1=0$, a situation which is easily realized for the
Ising magnet but not for a fluid; the containing walls 
 must always exert a non-zero surface field of some type. Usually the wall or substrate 
is assumed to exert an effective potential on the fluid that is infinitely repulsive
at short distances and strongly attractive at large distances. Such potentials give rise
to pronounced peaks in the density profile of the fluid  near the substrate leading to strongly 
positive adsorption and in  magnetic language this corresponds to $h_1\gg 0$.
On the other hand, the presence of  a substrate should decrease the net fluid-fluid
attraction between  an atom and its nearest neighbors below
the corresponding  bulk value so that $c>0$. Thus fluids  are expected to lie in the
universality  class of the normal transition with fixed point $h_1=\infty$ and $c=\infty$.
Since the shapes  of the order parameter  profiles  should be the same
  whether the symmetry  is broken by  
the external surface field, $h_1>0$, or spontaneously, $c<0$, the normal and extraordinary transitions
should be equivalent~\cite{bray:77:0,diehl:97:0}, provided the symmetry breaking fields 
are sufficiently (infinitely) strong.

In recent years much effort has been focused on understanding the crossover from the normal 
to the ordinary transition, i.e. the regime of weak surface fields.
The length scale associated  with the surface scaling field $h_1$ is 
$l_1\sim h_1^{-\nu/\Delta _1^{ord}}$, where $\nu$ is the bulk correlation length 
exponent  and $\Delta _1^{ord}$ is the surface gap exponent. For a semi-infinite Ising-like system
Ritschel and Czerner~\cite{uwe:96:0} 
showed that for small $h_1$ the order parameter profile can take a non-monotonic form, increasing with
distance $z$  from the surface for $z< l_1$ before decreasing for $z\gtrsim l_1$.
This is quite  different from the case of strong field adsorption  where the profile  decays monotonically,
as $z^{-\beta/\nu}$, at the critical point. Such behaviour is not predicted  by mean-field
treatments of Ising systems but was confirmed in several subsequent 
studies~\cite{ciach:97:0,uwe:97:0,czerner:97:0,czerner:97:1}; it is also found in exact
results  for the $d=2$ Ising model~\cite{bariev:88:0,konik:96:0,zamolodchikov:94:0}.  
Physically  $l_1$ corresponds  to the distance from the surface over
 which the order parameter responds linearly with field $h_1$~\cite{ciach:97:0}.

The existence of the critical length scale $l_1$ has important consequence for
{\it confined}  systems near bulk criticality. There are three relevant  lengths: $l_1$, the distance $D$ 
between the two confining surfaces which both exert fields $h_1$ and the bulk correlation
length $\xi _b$. Sufficiently close to bulk criticality one can investigate the
regime where $D$ is comparable to $l_1$ with $D, l_1\ll \xi _b$.
In a recent study~\cite{maciolek:99:0} of the critical $d=2$ Ising film order
parameter profiles with two symmetric maxima at $z\sim l_1$ and $z\sim D- l_1$ 
were obtained for weak fields $h_1$.
For even weaker fields  when $l_1\sim D$ the maxima merge into a single 
one at midpoint, $D/2$.
That study provides some of the motivation for the present  one where we investigate  the density profiles 
 and the Gibbs adsorption  for a simple (Lennard-Jones) fluid at bulk  criticality
that is confined by two planar walls separated by a distance $D$. We enquire
whether phenomena analogous  to these found for Ising systems subject to weak
surface fields are found in fluids.

As implied earlier, it is not at all obvious that one can  achieve the analogue for a fluid of 
surface magnetic field $h_1\to 0$. Thus it is not clear, a priori, that a physical
length scale $l_1$ should exist. On the other hand, one knows that for a purely
repulsive wall complete drying with adsorption $\Gamma\to-\infty$
 (wetting by gas) occurs below the 
critical temperature $T_c$ whereas for a strongly attractive  wall complete wetting  by liquid occurs, $\Gamma\to\infty$,
so that one might imagine a scenario whereby for {\it weak} wall-fluid attraction the adsorption
$\Gamma$ is close to zero. Moreover, if $\Gamma\approx 0$ for all states  on the critical isochore, 
$\rho=\rho _c$, $T\ge T_c$, one might argue that the fluid is exhibiting one of the
features of the Ising magnet subject to a vanishing $h_1$. In other words, by suitably tuning the
wall-fluid potential can one mimic  the situation $h_1=0$, i.e. the ordinary transition?
Of course, finding a wall-fluid potential that corresponds
{\it precisely} to $h_1=0$, i.e., one that generates precisely
the same features in the profiles and, therefore, the
adsorption for all $T$, seems very unlikely - given the lack of Ising symmetry in any 'real'
fluid.
However, if one does find  a wall-fluid potential that is 'neutral', giving
rise to $\Gamma \sim 0$, then one might also expect
 to see phenomena associated with small $h_1$, i.e. the  mesoscopic 
length scale
$l_1$. The existence of such phenomena would imply that the ordinary transition
is not unique to systems with Ising symmetry and would lend support
to the general idea of fluid-magnet universality.

We are not aware  of any published proof that a fluid system cannot undergo the ordinary transition~\cite{upton}.
The only rigorous result for fluids~\cite{upton:98:0} is for a planar hard wall where an
exact sum rule analysis shows that the surface critical behaviour lies  in the universality class of the normal 
transition and the surface critical exponents can be expressed in terms of bulk critical 
exponents, as predicted originally in Ref.~\cite{bray:77:0}.

What is the experimental situation? To the best of our knowledge
there is no experiment that probes critical adsorption in the weak surface
field regime for a one-component fluid~\cite{kiselev:00:0}.
For binary fluid  mixtures  there is strong evidence  from (i) reflectivity
studies of critical carbon disulphide-nitromethane mixtures
adsorbed at a silanated borosilicate glass  surface~\cite{franck:95:0}  
and (ii) ellipsometric studies  of an homologous series of critical
$n$ alkane - methyl formate mixtures~\cite{law:01:0} that the weak adsorption,
$h_1$ small limit, is achieved. In the second of these papers the authors fit 
their data to a universal scaling function which depends 
explicitly on $l_1$~\cite{law:01:0}. They argue that as the
 $n$ alkane  increases from 
$n=11$  to $n=14$ the preferentially  adsorbed component at the non-critical
gas/liquid mixture surface changes from  the $n$ alkane to methyl formate and that
 this implies an effective surface field $h_1$  which is small and which changes
sign as $n$ is increased. More specifically they associated $h_1$ with the 
difference  between the gas-liquid surface energies  of the $n$ alkane and
methyl formate in the mixture. In practice this was taken to be the difference 
between gas-liquid surface tensions of pure $n$ alkane and 
pure methyl formate~\cite{law:01:0}.
If  $h_1$ determines  whether a particular component is preferentially adsorbed at the interface 
it is certainly  feasible that one can have  $h_1=0$, at least for a certain range of temperatures. However, it is still
not clear that a given $n$ alkane-methyl formate mixture corresponds to a system undergoing an ordinary transition.
An earlier reanalysis of surface tension data for isobutyric-acid-water mixtures suggests that  the 
scaling properties  are those of a system  with a weak surface field, interpreted as the field that
determines the preferential adsorption of one component at 
 the non-critical interface \cite{fenzl:93:0}. ~The author also makes the important observation 
that similar weak-field characteristics should be found at the  non-critical interfaces
of any binary liquid mixture undergoing a wetting transition close 
 to the critical endpoint~\cite{fenzl:93:0,law:01:1}.
For a one-component fluid adsorbed at a solid substrate this situation corresponds 
to the wetting transition being close to the gas-liquid critical point, a scenario 
which is expected if the wall-fluid attraction  is weak.
Crossover between strong and weak-field  critical adsorption has been observed  
in very recent  X ray diffraction studies of 
sublattice order in epitaxial FeCo films grown on an MgO substrate~\cite{nickel:02:0};
FeCo undergoes a continuous order-disorder transition from the B2 phase (CsCl structure)
to the A2 phase (bcc structure) and the substrate favors (weak) 
critical adsorption of the B2 phase for $T\gtrsim T_c$.

Our paper is arranged as follows: in Sec.~\ref{sec:Is} we summarize results for the weak surface field behaviour 
of Ising systems in the semi-infinite lattice and in film geometry.
 The magnetization profiles obtained for the latter
 are directly relevant  to our
study of a confined fluid. Sec.~\ref{sec:sim} reports
 the results of Monte Carlo
simulations of the density profiles and adsorption of
 the truncated Lennard Jones fluid confined
between two planar walls that exert a standard (10-4)
  wall-fluid potential. Results
are presented at the bulk critical point, and  for some states on the critical isochore, for various values
of a parameter $f$ that multiplies the (10-4) potential. By varying $f$ we tune the strength of wall-fluid interactions relative 
to fluid-fluid interactions. $f$ plays a role similar to that of $h_1$  for the Ising system and for
certain values  of this parameter we find the system is 'neutral'.
This is confirmed by studies of capillary condensation below $T_c$.
In Sec.~\ref{sec:dft} we describe the results of density functional calculations of the density profiles for the same model system as studied
by simulation. By comparing the  two sets of results we can assess the role of true surface criticality 
since the density functional approach is a mean-field one. We conclude
 in Sec.~\ref{sec:concs} with a discussion and  summary.

\section{Summary of results for Ising systems} 
\label{sec:Is}  
\subsection{Semi-infinite lattice} 
\label{subsec:semin} 
We consider an Ising magnet  on a semi-infinite lattice with 
  nearest-neighbour pairs of spins  interacting (in  bulk)  with
 the exchange coupling
 constant $J$. The influence of the surface on the  system is usually
 taken into account by introducing  a modified coupling between spins 
in the surface layer, $J_1$, and an (external)
 surface  magnetic  field $H_1\equiv Jh_1$ imposed
 on these boundary spins~\cite{binder:83:0,diehl:86:0}. 
For vanishing bulk magnetic field, $H\equiv Jh=0$,  
the Hamiltonian of the model reads 
\begin{equation} 
\label{eq:IsHam}
 {\cal H}=-J\sum _{<ij>\in V}\sigma _i\sigma _j-J_1\sum _{<ij>\in\delta V}\sigma _i\sigma _j-H_1\sum _{i\in\delta V}\sigma _i 
\end{equation}
 where $V$ denotes the volume of the lattice  and $\delta V$
 the surface layer. In the absence of the surface field  the
 leading critical behavior of  various surface quantities depends
 on the ratio $J/J_1$ and is described by  three different
 surface universality classes~\cite{binder:83:0,diehl:86:0}. 
 At the bulk critical point, $H=0$, $~T=T_c$, and at the special
 value $J_{1c}\simeq 1.50J$ the tendency to order near the
 surface is unchanged with respect to the bulk. This situation  corresponds
 to the  surface universality class of the "special transition".
 The condition $J_1<J_{1c}$ represents the  universality class of
the  "ordinary transition" where the surface has a reduced tendency
 to order but becomes passively ordered at the bulk phase transition. 
 For $J_1>J_{1c}$  spontaneous symmetry breaking also occurs above
 the critical point and on reducing  $T$  the bulk orders  in the presence of an already
 ordered surface. This situation corresponds to the "extraordinary transition".
 In the renormalization-group analysis $J_{1c}$ is an unstable fixed point,
 whereas for any starting value   $J_1<J_{1c}$ the surface coupling is 
 driven to the stable fixed point $J_1=0$ by successive renormalization-group
 transformations. On the other hand, for $J_1>J_{1c}$ the coupling  is driven to the stable fixed point
  $J_1=\infty$.   In the framework of continuum field theory, appropriate
 to the  universality class of the Ising model,   the standard
 $\phi ^4$ Ginzburg-Landau model is augmented to include the surface by
adding the surface contribution: 
\begin{equation} 
\label{eq:GLH}
 {\cal H}_s(\phi)=\frac{1}{2}c\phi ^2-h_1\phi, 
\end{equation}
 where the parameter $c$ can be related to the surface
 enhancement of the spin-spin coupling constant in the lattice
 models~\cite{diehl:86:0}.  The leading critical behavior  of a 
semi-infinite system with an ordinary or an extraordinary surface 
transition  is described by  stable renormalization-group fixed-point values 
 $c=+\infty$ and $c=-\infty$, respectively. The fixed point for
 the special transition is at $c=c^*$.
So far we have not included  the effect of the surface field $h_1$ on the
 critical system.  The presence of the surface field  explicitly breaks the
 order-parameter (OP) symmetry  at the surface inducing ordering
 in the surface layer for $T \geq T_c$. It  has been
 argued~\cite{bray:77:0,diehl:97:0} that the  extraordinary transition,
 i.e. $h_1=0, J_1>J_{1c}$ ($c<c^*$), is equivalent  to the case with
 $h_1\ne 0$ but with de-enhanced surface couplings: $J_1<J_{1c}$
 $(c>c^*)$.
 The latter case corresponds to the so called "normal" transition
 described by the fixed point $h_1=\infty$~\cite{diehl:97:0}.
The term 'normal' is used because the 
 situation with $h_1\ne 0$ and $J_1<J_{1c}$ $(c>c^*)$ should 
 be generic  for simple fluids or fluid binary liquid mixtures bounded
 by a substrate or  wall.
 Most substrates  will attract molecules in a 
one component fluid giving rise to positive adsorption. Similarly for a binary mixture 
the substrate will favor one species of the fluid giving rise to preferential adsorption of that species.
Both situations correspond to $h_1\ne 0$.
Moreover for   fluids near walls the number of nearest-neighbor fluid 'bonds' 
is reduced with respect to the bulk  which implies $J_1<J_{1c}$ or $c>c^*$.

A topic of considerable recent interest  concerns
the crossover region between the limiting values $h_1=\infty$ and $h_1=0$.
  At the normal transition the
 OP (magnetization) profile $m(z)$ decays monotonically from $m_1=1$
(all spins aligned) at the surface $z=0$ to the bulk equilibrium value which is zero for $T\ge T_c$, 
 the bulk critical temperature and is nonzero for $T<T_c$.
 At the critical point the decay is described by a universal power
 law $m\sim z^{-\beta/\nu}$, where  $\beta$ and $\nu$  are the usual bulk exponents,
 for distances large compared to all length scales.
 In $d=3$ the Ising model  exponents give  
 $\beta/\nu = 0.518(7)$~\cite{ferrenberg:91:0}.
 By contrast, for $h_1=0$ and $T\ge T_c$, at the
 ordinary transition, the  magnetization profile is zero
 at any distance from the wall since the OP symmetry  is not broken
 through  the whole system~\cite{footnote}.  It is of considerable interest
to enquire about the profile for small but non-zero  
 $h_1$, i.e. in  the crossover regime between the ordinary
 and normal transitions.    Mean-field treatments  do not predict
 any interesting behaviour of the magnetization profile; $m(z)$
 starts from a  non-zero value  $m_1$ and then decays   monotonically  to
 zero~\cite{binder:83:0,bray:77:0,lubensky:75:0,peliti:83:0}. 
However, as has been  shown~\cite{uwe:96:0} recently, for small $h_1$ 
 close to the ordinary transition  the critical fluctuations
 may lead to  non-monotonic profiles. Near the surface $m(z)$ 
 may increase  steeply with $z$  to values much larger than $m_1$.
 Near bulk criticality  the range over which this rise occurs depends  
 on $h_1$ through the length scale induced by this field. 
In an Ising system near the ordinary transition, i.e. for  $J\ll J_{1c}$,
 there is a single relevant  surface scaling field $h_1$~\cite{diehl:86:0} 
and the length scale induced by this  field is $l_1\sim h_1^{-\nu/\Delta _1^{ord}}$,
 where  $\Delta_1^{ord}$ is the surface gap exponent~\cite{kom:1}.
 At the bulk critical point  the magnetization is
 given by 
\begin{equation} 
\label{eq:mag}
 m(z)\sim h_1z^{\kappa}~~~~~~$for$~~~z\lesssim l_1
\end{equation}
with $\kappa=(\Delta_1^{ord}-\beta)/\nu$  ~\cite{uwe:96:0,uwe:97:0}. 
 In the $d=3$ Ising model $\Delta_1^{ord}>\beta$ and $\kappa \simeq 0.21 $
so $m(z)$ {\em increases} with $z$ near the surface.
 For $ z\sim l_1$ the profile
 has  a maximum, and only for $z\gtrsim l_1$ does crossover to the 'normal'
 decay $\sim z^{-\beta/\nu}$ take place. For weak surface fields  $l_1$
 can be large ( mesoscopic ) and thus the increase of the  magnetization profile can
 occur over a  mesoscopic distance. In turn, this  may influence
 the behavior of the critical
  adsorption slightly above $T_c$~\cite{ciach:97:0,franck:95:0,law:01:0}. 
  The short-distance behavior of the magnetization profile,
 Eq.~(\ref{eq:mag}),
 follows from the critical-point   scaling of $m(z)$ and the
(continuity) assumption that $m(z\to 0)\sim m_1$.  Since at the ordinary
 transition the surface layer remains paramagnetic at $T=T_c$ it should
 respond linearly to a weak  surface field $h_1$~\cite{bray:77:0}.
 The same should hold in the immediate neighbourhood of  a surface,
 thus  $m(z)\sim m_1\sim h_1$ for  $z\to 0, h_1\to 0$.  
 Assuming linear response for the scaling function leads to 
 (\ref{eq:mag})~\cite{uwe:96:0}.  A further heuristic explanation of the
 short-distance growth of the OP at the ordinary transition was
 given in Ref.~\cite{uwe:96:0,uwe:97:0} on the basis of the behaviour of 
correlations in the near-surface  region. Using scaling analysis 
 and arguments taken from short-distance  expansions it was argued
 that there exists  an {\it effective} parallel correlation length
 $\xi _{\parallel}(z)$ which marks the crossover of the  parallel
 correlation function from  "bulk"  to "surface"  behavior. At $T=T_c$
 the correlations decay algebraically.  The decay in planes parallel
 to the surface is described by $r^{-d+2-\eta _s}$,  where $r$ is the
 distance within the plane parallel to the surface  and the exponent
 $\eta _s$ depends on the distance $z$ from the surface.  For $z<l_1$,
  in the near-surface region, $\eta_s$ assumes its bulk 
 value $\eta$ (slow decay) if $r\ll z$ whereas for $r \gg z$,
 $\eta_s=\eta^{ord}_{\parallel}>\eta$ (fast decay), so we can infer 
the effective range of correlations 
 $\xi_{\parallel}\sim z$. Now  assume a weak surface field is imposed   at bulk criticality
 on the surface with $J_1 \ll J_{1c}$.
 The magnetization induced in response  at a distance $z$ from
 the surface   should be proportional to $h_1$,  to the correlation
 function in the  perpendicular direction
describing the decay of the order, 
 $<m(0)m(z)>\sim z^{-d+2-\eta_{\perp}^{ord}}$, 
and to the correlated area in  the plane parallel to the surface,
 $\xi_{\parallel}^{d-1}$, which can be  influenced by a single surface spin.
 Together these terms give 
\begin{equation} 
\label{eq:uwe}
 m(z)\sim h_1 <m(0)m(z)> \xi_{\parallel}^{d-1}\sim h_1 z^{1-\eta_{\perp}^{ord}}~~~~~~$for$~~~~~~z<l_1
\end{equation}
 Scaling relations \cite{uwe:97:0}
 give in turn $1-\eta_{\perp}^{ord}=\kappa$  and we recover (\ref{eq:mag}). Many of the above predictions
 were confirmed  by Monte-Carlo simulations for the $d=3$
 Ising model ~\cite{czerner:97:0}  and by detailed
 renormalization group studies of a $\phi ^4$ model~\cite{uwe:97:0}.
 Exact results~\cite{bariev:88:0,konik:96:0,zamolodchikov:94:0} and Monte Carlo simulations~\cite{czerner:97:1}
for the  $d=2$ Ising model yield  magnetization profiles that also 
 increase  near the surface for the case of a weak surface field. However, in 
$d=2$ the simple power law (\ref{eq:mag})  is modified by a logarithm, i.e. 
$m(z)\sim h_1z^{\kappa}\log (h_1z)$, where $\kappa=3/8$. 
Note that in mean-field theory $\Delta_1^{ord}=\beta=1/2$ so that $\kappa=0$ and the magnetisation profile decays monotonically as stated above.

\subsection{Ising films} 
\label{subsec:films}

In this subsection we consider film geometry, i.e. a lattice of finite width  $D$ in the $z$
direction but infinite in the other directions.
 The critical properties of such a confined system should be
 particularly  sensitive to the value of the (identical) surface fields when the  
 length scale $l_1$ becomes comparable to or larger  than the width $D$. 
 Recent studies of critical $d=2$ Ising films
 of  finite width $D$  subject to identical surface fields $h_1=h_D$
 using exact transfer-matrix diagonalization and  density-matrix
 renormalization-group (DMRG) methods fully confirmed 
 this expectation~\cite{maciolek:99:0}. For weak surface fields  
 such that $l_1< D$, magnetization profiles at $T=T_c,~ h=0$
   have two symmetric maxima shifted away from the walls and lying at
 $z\sim l_1$ and $z\sim D-l_1$ (see Fig.\ref{fig:Ispr}). 
 In this regime of $h_1$ the surface magnetization $m_1=m_D$ decreases  fairly
 rapidly as the value of $h_1$ decreases, whereas the value of the
 magnetization at the midpoint of the film, $m(D/2)$,  changes only very slightly.
 The two separate maxima  merge into one, broad maximum located  at
 the middle of the film for smaller $h_1$ where $l_1 \sim D$. For even smaller $h_1$,
$l_1>D$ and the profiles  are
 nearly flat with the magnetization much lower near the walls than 
 in the central part of the system. In this regime  the
 magnetization both at the surface and near the midpoint, in fact
through  the whole profile, decreases  as $\sim h_1\ln h_1$ (the logarithmic
 factor is specific  to the  $d=2$ Ising model).  This result  agrees
 with the physical interpretation of $l_1$ as a length scale 
 up to which the OP profile responds linearly (up to the logarithmic factor)
 to the surface field. Note that the change in the shape of the magnetization
 profile from one with  two maxima to  one with a single maximum
 at the center of a film can be achieved either by decreasing $h_1$ at
 fixed $D$ or by decreasing $D$ at fixed $h_1$. This change is accompanied
 by a rich variation of the  solvation force~\cite{maciolek:99:0}.
The latter is  an important thermodynamic
 quantity characterizing a confined system~\cite{evans:90:0}. For a fluid  the
 solvation force  is  the excess pressure over the bulk 
arising from  confinement by walls; it can be measured
 by the surface force apparatus. For the Ising system the solvation force
$=-(\partial f^*/\partial D)_{T,h_1,h}$, where  $f^*(D)$ is the finite-size contribution to the free energy.
 The critical-point scaling function
 of the solvation force, i.e. at $T=T_c$ and $h=0$,  has a pronounced maximum
  near $l_1=D$ (see Fig.6 of Ref.\cite{maciolek:99:0}).
       
\section{Simulation studies of the  Lennard-Jones fluid in a slit pore} 
\label{sec:sim}  

In this section we describe the results of a Monte Carlo study of a three dimensional system, 
namely the truncated Lennard-Jones
fluid, confined between two planar walls separated by a distance $D$. By decreasing the strength $f$
 of the wall-fluid interactions while keeping the fluid interparticle potential  constant we examine the effects of 
decreased wall-fluid attraction on the shapes of the density profiles and on the form of  the adsorption in the critical fluid.
Decreasing $f$ should, in some sense, correspond to reducing $h_1$ in the Ising system discussed in the
previous section.
\subsection{Computational details} 
\label{sec:comp}  
We have performed Monte Carlo simulations of the Lennard-Jones (LJ) fluid 
having interparticle interactions of the form:  
\begin{equation}  
U_{LJ}(r)=4\epsilon\left[\left (\frac{\sigma}{r}\right)^{12}-\left(\frac{\sigma}{r}\right)^{6}\right ]\;.  
\label{eq:lj}  
\end{equation}  
Here $\epsilon$ measures the well depth of the potential, while 
 $\sigma$ sets the length scale.  
The potential was truncated at a radius $r_c=2.5\sigma$ and  left unshifted. 
No corrections were applied to account for effects of the truncation. 
  The simulations were performed within the grand canonical (constant-$\mu V T$) 
ensemble \cite{FRENKEL}, permitting fluctuations in the total particle number $N$.
 In order to mimic a slit pore geometry, a cuboidal simulation cell of dimensions 
$L_x\times L_y\times D$ (with $L_x=L_y=L$) was employed. 
Structureless planar walls were imposed in the planes $z=0$ and $z=D$, 
while periodic boundary conditions were applied to the cell boundaries 
in the $x$ and $y$ directions parallel to the  walls.   
Fluid particles were assumed to interact with a single planar wall via a 
long range potential having the form:  
\begin{equation} 
\label{eq:extlr}
U_w(z)=4\epsilon f\left[ \frac{2}{5}\left(\frac{\sigma}{z}\right)^{10}-\left(\frac{\sigma}{z}\right)^{4}\right ]\;\;
\end{equation}
 where $f$ is a parameter that tunes the strength of the
 fluid-wall interactions relative to those of the fluid interparticle
 interactions. Decreasing $f$ simply reduces the depth of the minimum.
The total wall-fluid potential is $U_w(z)+U_w(D-z)$.
 We note that $U_w(z)$ models a wall comprising a single
 plane of LJ particles \cite{ISRAEL}. No potential truncation was
 applied to $U_w(z)$ since the wall potentials decay considerably less 
rapidly with increasing separation than the LJ interparticle potential
 of eq~(\ref{eq:lj}). Further details of the simulation procedure are given in an earlier study~\cite{maciolek:99:1}.
  In the course of the simulations the one-dimensional
 density profile was accumulated:  
\begin{equation} 
\rho(z)\equiv\int_0^L \rho({\bf r})dxdy  \;\;, 
\end{equation}
 representing the configurationally averaged local number
 density at a given distance $z$. Measured forms of this profile for a
 range of  wall strengths $f$ at the liquid-gas critical point
and along the critical isochore  are presented in the following subsection.  
\subsection{Density profiles at the critical point} 
\label{sec:pore} 
 The bulk critical point parameters of the truncated LJ fluid
 have previously been determined from an accurate finite-size scaling
 study  \cite{WILDING95}. They are (in standard reduced LJ units
 \cite{FRENKEL}), $T_c=1.1876(3), \mu_c/k_BT_c=-2.778(2)$. 
Using these parameter values we have obtained the density
 profiles $\rho(z)$ for a range of wall strengths $f$.
 The results for a system of size $L=15\sigma,D=40\sigma$ are
 shown in Fig.~\ref{fig:critprofs}(a). A magnified plot of the region around the critical density $\rho_c=0.3197(4)$ (horizontal line) is shown in Fig.~\ref{fig:critprofs}(b). 
Examination of the profiles of Fig.~\ref{fig:critprofs} reveals 
three qualitatively distinct regimes of behaviour as the wall
 strength $f$ is varied. We discuss them in turn.   For the smallest
 values of $f$ studied, the density near the wall lies well below its
 critical value and the profile exhibits little structure. Moving away
 from the wall, however, the density increases, reaching a maximum in the
 slit middle, although it at no stage attains the bulk critical value.
 The overall shape of the profile is convex upwards with respect to
 the critical density.  For the largest values of $f$ studied,
 the density near the wall is much higher than the critical density 
and large packing effects are evident in the profile. As the distance 
from the wall increases, the density falls progressively, although it
 does not reach the bulk value. The profile is convex downwards with
 respect to the critical density.  For intermediate values of $f$, 
 the packing-induced density oscillations near the wall span the critical
 density, while further away from the wall the magnitude of the profile
 curvature is generally less than in the large or small $f$ limits.
 A magnified view of the region around the critical density is given
 in Fig.~\ref{fig:critprofs}(b). Strikingly there exist some  $f$
 values in this regime for which $\rho(z)$ exceeds $\rho_c$,
 but the profile is {\em concave  upwards} with respect to
 the critical density.  Thus $\rho(z)$ exceeds $\rho_c$
 and {\em increases} with increasing distance from the wall.
 As $f$ increases, however, one sees (Fig.~\ref{fig:critprofs}(c))
 the formation of a  structure with two smooth maxima in the profile,
 the maxima being some distance from the walls. On further increasing
 $f$, the maxima become more pronounced  and their positions move closer to the walls.
Clearly these results show very similar features to the magnetisation profiles shown in Fig.\ref{fig:Ispr}. 
   It is also instructive to observe the effect of changing the slit width on
 the double maxima  profile of Fig.~\ref{fig:critprofs}(c).
 The comparison with a profile for a slit of width $D=20\sigma$
 is made in Fig.~\ref{fig:fss},
 from which one sees that the double maxima  profile is replaced by one with a single maximum
in the narrower slit. Once again this is  similar 
to what is found for the Ising system (Sec.~\ref{sec:Is}).

\subsection{Properties of the Neutral Wall System}
\label{subsec:nw}

The results of the last subsection suggest that for certain choices of the strength parameter
$f$ the confined Lennard-Jones fluid
is behaving similarly to an Ising film with $h_1\approx 0$. Density profiles for $f\lesssim 0.653$ 
are of particular interest since it is in this regime  of $f$ that there is a crossover from a 
 maximum to a minimum in the profile at the midpoint.

One of the signatures of the ordinary transition (at a single surface) is that the magnetisation profile remains constant,
 $m_l=0, l=1,2,\cdots$ for all $T\ge T_c$ and $h=0$. The analogous situation for  a confined fluid  should be $\rho (z)=\rho _c$, 
the critical density, except for distances close to the wall where packing constraints induce oscillations in the profile, for
states  on the critical isochore: $\rho _{bulk}(\mu,T)=\rho _c, T\ge T_c$.
This implies, in turn,  the adsorption $\Gamma=\int _0^D(\rho(z)-\rho _c)dz$ should 
be constant (and close to zero) along the critical isotherm.   

We varied $f$ in an attempt to find a 'neutral' wall where such properties were achieved. Figs~\ref{fig:neupr}(a)
and \ref{fig:neupr}(b) display the density profiles on the critical isochore for $f=0.644$, $D=40\sigma$ and $20\sigma$, respectively. 
In each case the density profile is  almost flat throughout the slit except for near the walls.
Fig.\ref{fig:ads} shows a plot of the adsorption $\Gamma \sigma ^2$ versus $\tau=(T-T_c)/T_c$ for $D=20\sigma $ and a selection of values of $f$. 
We observe that $\Gamma $ is temperature independent for the 'neutral' wall with $f=0.644$.
Increasing $f$ slightly leads to positive adsorption near $T_c$ whereas decreasing $f$ below $0.644$ leads to more pronounced negative adsorption which is $T$ dependent.

Note that the simulations were performed at chemical  potentials $\mu$ measured along the critical isochore of the bulk fluid.
These were determined, using periodic boundary conditions, in Ref.~\cite{maciolek:99:1}.

Further evidence that $f=0.644$ corresponds to a 'neutral' wall is furnished  by considering the confined fluid  {\it below} $T_c$. 
For strongly attractive walls, favouring the liquid phase, we expect
 liquid-gas coexistence to occur at a chemical potential $\mu$ which lies below the value $\mu_{cx}$ for bulk coexistence at the same
 temperature~\cite{evans:90:0} whereas for purely repulsive walls, favouring gas, coexistence should occur at a higher value than in bulk.
We determined the line of capillary coexistence , where liquid and gas coexist, in the slit with $D=20\sigma$. 
For $-0.1<\tau <0$ and $f=0.644$ we  found this line was hardly  shifted from the bulk coexistence  line - see Fig.~\ref{fig:cap}. 
The greatest deviation, at $\tau=-0.1$, was about $0.1\%$, with condensation in the slit ocurring for a (slightly)  higher value of $\mu $, 
implying the walls with $f=0.644$ favor (slightly)  the gas phase, i.e. the contact angle is (slightly) larger than $\pi/2$ at this temperature~\cite{evans:90:0}. The inset to Fig.~\ref{fig:cap} displays the difference between the chemical potential $\mu$, for capillary coexistence, and the corresponding value
$\mu_{cx}$, for bulk coexistence, for three values of $f$. One observes that for $f=0.71$, $\mu-\mu_{cx}$ remains substantially more negative 
than the result for $f=0.644$ whereas for $f=0.59$, corresponding to a weaker
wall attraction, $\mu-\mu_{cx}$ is positive at all temperatures in the range
considered. These results suggest that a value of $f$ slightly lower than
 $0.644$ would yield capillary coexistence very close to that in bulk, i.e. would correspond in magnetic language to a surface field $h_1=0$ as regards the phase behaviour below $T_c$.

\section{Density-functional theory studies of
 the Lennard-Jones fluid in a slit pore} 
\label{sec:dft}

Having established that computer simulations of the critical  Lennard-Jones
fluid do find features in the density profile that mimic those found for the magnetisation  profile in systems with weak surface fields $h_1$, we now enquire 
whether any of these features are also found  in a mean-field treatment
of the same confined fluid. The oscillations in the profiles 
 shown in Figs 2-4, arise from packing effects near the walls and these should
certainly be captured by a suitable  density functional theory  (DFT) approach.
Recall, however, the double maxima, separated by a shallow minimum, is a feature which reflects true surface criticality in the case of a magnetic system (Sec.~\ref{sec:Is}). Does the presence of a similar feature in  the simulation results  for fluids also signal some manifestation of surface criticality or is the feature somehow reflecting the particular form of the wall-fluid  potentials?
By carrying out mean-field calculations we attempt to address these questions.

We consider the same system of particles as in  Sec.~\ref{sec:sim}, 
 i.e. the Lennard-Jones fluid with the pair potential
 given by ~(\ref{eq:lj}), but {\it not} truncated, and apply
  the DFT~\cite{evans:79:0} 
 to determine  the density profile at the bulk critical point.
 All the  functionals implemented here are of  mean-field type; 
bulk critical exponents  would take their mean-field values.
  We assume that the fluid is confined  between two identical
 parallel walls located at $z=0$ and $z=D$, and infinite in the  $x$
 and $y$ directions. As previously, the system is in contact 
with a reservoir of 
 fluid at  temperature $T_c$ and  critical chemical potential $\mu _c$.
 The grand potential of the inhomogeneous  system is a  functional of 
the one-body  density
 distribution $\rho({\bf r})$~\cite{evans:79:0}: 
\begin{equation} 
\label{eq:gpotfun1} 
\Omega[\rho]= {\cal F}[\rho]-\int d{\bf r}(\mu-V({\bf r}))\rho({\bf r}), 
\end{equation}
 where  $V({\bf r})\equiv V(z)$ is the total wall-fluid  external potential 
and the equilibrium density profile  $\rho({\bf r})\equiv \rho(z)$
 corresponds  
to the minimum of $\Omega[\rho]$.
  The simplest form for  $\Omega[\rho]$ which should be appropriate near 
criticality  is based on the square gradient approximation  
 to the intrinsic Helmholtz free energy functional  ${\cal F}[\rho]$ with  the 
  wall-fluid contribution modeled    by a term $\Phi_s$ which depends 
only on the  fluid density at contact i.e. 
 $\Phi_s=\frac{c}{2}(\rho^2(0)+\rho^2(D)) -\varepsilon _w(\rho(0)+\rho(D))$,
where $c$ and $\varepsilon _w$ are constants characterizing the effects of the wall. 
Although functionals of this type   cannot incorporate short-ranged correlations arising from packing  and 
 hence cannot acount for the oscillatory behavior of the density profile  
 near the walls~\cite{evans:90:0}, they were  succesfully employed 
  in   studies  of finite size effects on the  critical adsorption in large pores, 
as this phenomenon is dominated by the  slow decay of the  profile far
 from the walls~\cite{marini:88:0,maciolek:99:1}.  However,  
the density profiles obtained from  this approximation  
 are always monotonic; depending on the value of 
 $c$ and $\varepsilon _w $, $\rho(z)$ is either monotonically
 increasing or monotonically decreasing. This observation 
follows straightforwardly from 
the graphical construction for  the solution of the Euler-Lagrange (differential) equation ~\cite{marini:88:0,evans:85:0}
and remains valid for the local functional approach employed by Borjan and Upton~\cite{borjan:98:0}
in their studies of a  critical confined system. 
Thus it is necessary to  consider other  approximations  for the intrinsic free energy 
functional~\cite{evans:79:0,evans:92:0},  in which  ${\cal F}[\rho]$ is divided into
  contributions to  the free energy  functional  arising from  repulsive forces, 
 ${\cal F}_{r}[\rho]$, and from the attractive part  of the pairwise potential; 
 the latter is  treated in mean-field or random-phase approximation. 
Using hard-spheres (HS) as the reference fluid one constructs a Van der Waals approximation:
 \begin{equation} 
\label{eq:vdw} 
{\cal F}[\rho]={\cal F}_{HS}[\rho]+\frac{1}{2}\int d{\bf r}_1\int d{\bf r}_2\rho({\bf r}_1)\rho({\bf r}_2)\phi_{att}(r_{12}),
\end{equation} 
where $r_{12}=|{\bf r}_1-{\bf r}_2|$.
 The attractive part of the  pairwise potential is not specified uniquely. Here we employ the prescription:
 \begin{eqnarray} 
\label{eq:div}
 \phi_{att}&=&U_{LJ}(r_0)~~~~~~~~~~~~  r\le r_0 \nonumber \\
           &=&U_{LJ}(r)  ~~~~~~~~~~~~  r>r_0, 
\end{eqnarray}  
where $r_0=2^{1/6}\sigma $ is the minimum  of the  LJ potential and 
$U_{LJ}(r_0)=-\epsilon$.
In this approximation the Helmholtz free energy density of a uniform fluid has the  form:
\begin{equation}
\label{eq:free}
f(\rho)=f_{HS}(\rho)-\frac{\rho^2}{2}\alpha,
\end{equation}
with $\alpha=-\int d{\bf r}\phi _{att}(r) >0$.
For simplicity we set the hard sphere diameter $d$ equal to $\sigma $.
We describe results of calculations using 
 both the weighted density approximation (WDA)
and the local density approximation (LDA)  for ${\cal F}_{HS}[\rho]$~\cite{evans:92:0}.
 
\subsection{Results from WDA} 
\label{subsec:wda}  
In order to incorporate   correctly the main features of the  short-range structure 
 of a fluid near the walls, 
 we employ a simple non-local approximation for ${\cal F}_{HS}[\rho]$ based on 
the weighted-density  approximation~\cite{tarazona:84:0,tar:84:1,tarazona:85:0}: 
\begin{equation} 
\label{eq:nonloc}
 {\cal F}_{HS}[\rho]={\cal F}_{id}[\rho]+\int d{\bf r}\rho({\bf r})\psi_{ex}(\bar{\rho}({\bf r})),
\end{equation}
 where the ideal gas contribution is
\begin{equation}
\label{eq:fid}
{\cal F}_{id}[\rho]=\int d{\bf r}\rho({\bf r})[\ln (\Lambda^3\rho({\bf r}))-1].
\end{equation}
$\Lambda$ is  the thermal de Broglie wavelength.
$\psi_{ex}$ is the excess, over ideal, of the   free energy per particle  of the uniform fluid. In (\ref{eq:nonloc}) this quantity  is   calculated at 
some  weighted density $\bar{\rho}$, where
$\bar{\rho}$ is a non-local functional of the true density $\rho$ 
obtained by taking the weighted average  of $\rho (\bf{r})$ over a local 
volume reflecting the range of interatomic forces: 
\begin{equation} 
\label{eq:weight} 
{\bar \rho}({\bf r})=\int d{\bf r'}\rho ({\bf r'})w(|{\bf r}-{\bf r'}|). 
\end{equation} 
The precise definition 
of $w$ depends on the particular version of WDA;
the most sophisticated versions introduce weight(s) so that the second functional derivative of  ${\cal F}_{HS}[\rho]$
 reproduces
the two-particle direct correlation function  $c^{(2)}(r)$ of  a uniform
hard sphere fluid~\cite{evans:92:0}.
In this paper we assume  the simplest, density independent step function form 
for  $w(r)$, i.e.  $w(r)=3/(4\pi d ^3)\Theta (d-r)$, where $\Theta(x)$
 denotes the Heaviside step function. This choice   yields 
reasonable results for the density profiles near a  hard wall~\cite{tar:84:1}.
The excess free energy per atom is given by the Carnahan-Starling expression:

\begin{equation} 
\label{eq:freeen} \psi _{ex}(\rho)=k_BT\frac{\eta(4-3\eta)}{(1-\eta)^2}, 
\end{equation}
 where $\eta=\frac{1}{6}\pi \rho d ^3$ is the packing fraction.
 Minimizing $\Omega[\rho]$ in (\ref{eq:gpotfun1}) with the external potential
$V(z)$  yields an integral
equation for  the density profile of the  fluid:

\begin{eqnarray}
\label{eq:inteq}
\rho(z)& = & \rho _0\exp \{ -(1/k_BT)[\psi_{ex}(\bar{\rho}(z))+\int d{\bf r'}\rho(z')\psi'_{ex}(\bar{\rho}(z'))w(|{\bf r}-{\bf r'}|) \nonumber \\
       & + & \int d{\bf r'}\rho(z')\phi_{att}(|{\bf r}-{\bf r'}|)+V(z)]\} ,
\end{eqnarray}
where $\rho_0=\Lambda ^3\exp(\mu/k_BT)$.
  
We solve this equation numerically for two choices of the single wall potential $U_w(z)$:
(a) a long range potential with the same form used in simulations (\ref{eq:extlr})
and (b) a hard wall potential with  an attractive short-ranged tail:
\begin{eqnarray} 
\label{eq:extsr}
U_{ws}(z) &=&\infty ~~~~~~~~~~~~~~~~~~~~~~~~~~~~~~  z< 0 \nonumber \\
          &=&-\epsilon f\exp (-z/\sigma) ~~~~~~~~~~~~  z>0. 
\end{eqnarray}  
Once again the strength parameter $f$ determines the depth of the potential well.
The integral equation (\ref{eq:inteq}) was solved  
using a minimization  procedure~\cite{poniewierski:93:0} which, 
in principle, is always convergent;  the 
 commonly used iteration
is  much less efficient near bulk  criticality where profiles decay very slowly.
We minimimize the function
\begin{equation}
\label{eq:fmin}
u(\rho_1,\ldots,\rho _N)=\sum _{i=1}^{N}(\rho _i-\tilde{\rho}_i)^2,
\end{equation}
where $\rho _i=\rho (i)$ are densities calculated at $N$ points between $z=1$ and $z=D$ 
and are treated as independent  variables.
$\tilde{\rho}_i$ denotes the right-hand side of the integral equation (\ref{eq:inteq})
calculated at $z=z_i$; $\tilde{\rho}_i$ are functions of $\rho _i$, i.e.
$\tilde{\rho}_i=\tilde{\rho}_i(\rho_1,\ldots,\rho_N)$.
The minimum $u\approx 0$ corresponds to the solution of (\ref{eq:inteq}).
This minimum is determined  using the conjugate gradient method.
Calculations were performed for the slit pore of the same width as in the
simulations,  $D=40\sigma$, at the critical point of the uniform fluid.
The mean field free energy density (\ref{eq:free}) gives,  for our model,
 $k_BT_c/\epsilon=1.00617(2)$ and $ \rho _c\sigma ^3=0.24912(4)$.
To allow for  the expected oscillatory behavior of the 
 profile near the walls we 
 used a very fine mesh,  $N=400$.
As a  starting density profile we usually took  $\rho (z)=\rho _c$.

In Fig.\ref{fig:WDALR}(a)
we show  density profiles obtained for the long-ranged 
 wall-fluid 
potential (\ref{eq:extlr}) for  various strengths $f$ between 0 and 1.
The sequence of   profiles is
 very similar  to that 
obtained in simulations  (see Fig.~\ref{fig:critprofs}(a)), i.e. there
are three different types of  profile  shape depending on the value of $f$.
For the smallest values of $f$ studied ( between 0 and about 0.55),
 the profiles are convex upwards with respect to $\rho_c$ and exhibit
 little structure, 
whereas  for the largest values of $f$ studied (between about 0.7 and 1)
the  profiles 
are convex downwards with $\rho(z)>\rho_c$ and show very pronounced
oscillations near the walls. 
At intermediate values, i.e. as $f$ increases above  $\approx 0.55$ the profiles  become
flatter in the central part of the slit. They remain concave upwards
with respect to the critical density but   move as a whole
  towards greater values
 so that except near the walls,  $\rho (z) > \rho _c$- see
 Fig.~\ref{fig:WDALR}(b). 
A  difference between results from the WDA and those  from simulations 
 occurs when the two smooth maxima, separated by a shallow minimum, structure
forms in the profile. In the simulations these new maxima  are 
 more pronounced 
than in WDA and  are located somewhat  closer  to the walls.
It is instructive to compare the height and position
 of the new maxima with those  of the third peak in the density profile, arising from packing effects (layering) near the walls. In the WDA this third peak
is located at  $z\approx 3\sigma$. For $f\lesssim  0.55$ the peak is not pronounced indicating weak packing effects in the weakly attractive wall potential.
As $f$ is increased the third peak becomes more pronounced  and the value of the density  at this peak increases  rapidly. For $f\lesssim 0.63$ this density lies below the density at the midpoint, $\rho _{mid}$, and below the density 
at the new maxima (where these have developed). However for $f$  in the range 
 $0.63\lesssim f\lesssim 0.64$ the density in the region
of the third peak grows more rapidly than in the central portion of the density profile and for $f\gtrsim 0.64$  the density at the third peak is higher than at the new maxima and $\rho _{mid}$. The positions  of the new maxima move closer to the walls as $f$ increases and for $f\gtrsim 0.68$  these maxima cannot be discerned.
Note that the range of values of $f$ corresponding to the  three qualitatively
different  regimes of  behavior   of the profile 
are very close to those found in simulations. 

In Fig.~\ref{fig:ads1} we plot the adsorption $\Gamma\sigma^2$ versus $\tau$, 
for $D=40\sigma$, calculated along the critical isochore $\rho=\rho_c$, for 
several values of $f$.These results should be compared with those from simulation in Fig.~\ref{fig:ads} noting that the latter correspond to a {\it much}
 wider range of the strength parameter $f$. We should also note the difference
in vertical scale between these two figures; the adsorption is very weak
within the WDA for the narrow range of $f$ displayed in Fig.~\ref{fig:ads1}.
On the basis of this figure we might conclude  that a value
of $f$ in the range  $0.561\lesssim f\lesssim 0.564$
corresponds to a 'neutral' wall. In this range  $\Gamma$ changes from
monotonically increasing with $\tau$  to exhibiting
a shallow minimum. There is some value for which $\Gamma (\tau)$ is
flat and negative, c.f. Fig.~\ref{fig:ads}.

 Figure \ref{fig:WDASR}(a) presents a  selection of the profiles calculated for
 the short-ranged  wall-fluid potential  
(\ref{eq:extsr}) with  various  strengths  $f$  between 1 and 0.
There are generally fewer  oscillations near the walls
and the contact values  $\rho(0)=\rho(D)$ are large for large values of $f$
but away from the walls the shape of the profile
changes in a similar  way as
for  the long-ranged  $U_w(z)$  
as $f$ is increased.
The two smooth maxima  structure  forms at  much larger values of $f$ 
however, i.e., at about
0.8 and then shifts rapidly towards the walls, merging with oscillations
 - see Fig.\ref{fig:WDASR}(b). Such behaviour might reflect the narrower
attractive well in the wall-fluid potential. One requires a larger $f$ to acquire the same strength of attraction.  In the  range of $f$ we  studied there  
are no profiles characteristic of strongly adsorbing walls with 
 extreme packing effects.

\subsection{Results from LDA} 
\label{subsec:lda}  

In the local density approximation (LDA) the free energy functional of the 
hard-spheres has a form:
\begin{equation}
\label{eq:ldafe} 
{\cal F}_{HS}[\rho]\approx \int d{\bf r} f_{HS}(\rho({\bf r})),
\end{equation}
where $f_{HS}(\rho)$ is the Helmholtz free energy density of a uniform hard-sphere
fluid with  density $\rho$   \cite{evans:79:0,evans:92:0}.
The LDA does not incorporate the short-ranged correlations that characterize the structure of  dense 
liquids and hence it cannot describe oscillatory profiles arising from packing effects. Attractive 
interparticle forces  are still treated by the mean-field approximation given  in (\ref{eq:vdw}) and the free energy density of the uniform fluid  is still given by (\ref{eq:free}) so that the critical point remains the same as that given 
in the previous subsection.

Critical point  profiles for the slit of  width $D=40\sigma $
 obtained in the LDA for the long-ranged  wall-fluid potential (\ref{eq:extlr}) are shown in  Fig.~\ref{fig:LDA}(a).
The wall strengths $f$ are the same as in Figs.\ref{fig:WDASR}. Despite  the very crude  approximation,
  some weak structure remains in the profiles. 
The sequence of  profiles
is reminiscent in shape  to those  from Fig.~\ref{fig:WDASR}, i.e.
for  WDA with the short-ranged wall-fluid potential.
For  $f< 0.8$ the  shapes  are  convex upwards with  respect to the critical density
and, except for  the first peak near the walls, $\rho(z)$   lies below $\rho _c$.
Profiles for $ f\ge 0.8$  
 are similar in shape to these  obtained in 
simulation and in WDA  for intermediate  values of $f$ - see Fig.~\ref{fig:LDA}(b).
As was  the case for  WDA with the short-ranged wall-fluid potential,
the characteristic two smooth maxima structure is formed for larger values of $f$
than in simulations.  
 
\section{Discussion}
\label{sec:concs} 
\begin{figure}[h]  
\setlength{\epsfxsize}{7.2cm} 
\centerline{\mbox{\epsffile{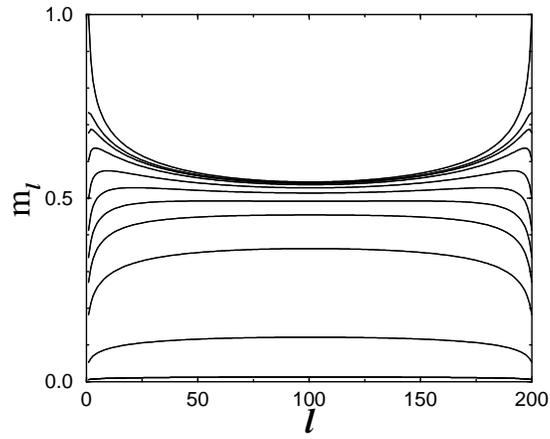}}}  
\caption{ Magnetization profiles $m_l$ of  the $d=2$ Ising film of  
width $D=200$ lattice spacings for $T=T_c$ and  zero bulk field and several values of the surface 
field $h_1=h_D$: from  top to  bottom profile: 
$h_1=10, 0.5, 0.4, 0.3, 0.2,0.14,0.1,0.07,0.04, 0.01, 0.001$. 
These results were obtained  using exact transfer-matrix 
diagonalization (adapted from Ref. 13). } 
\label{fig:Ispr} 
\end{figure}

\begin{figure}[h]  
\setlength{\epsfxsize}{7.2cm} 
\centerline{\mbox{\epsffile{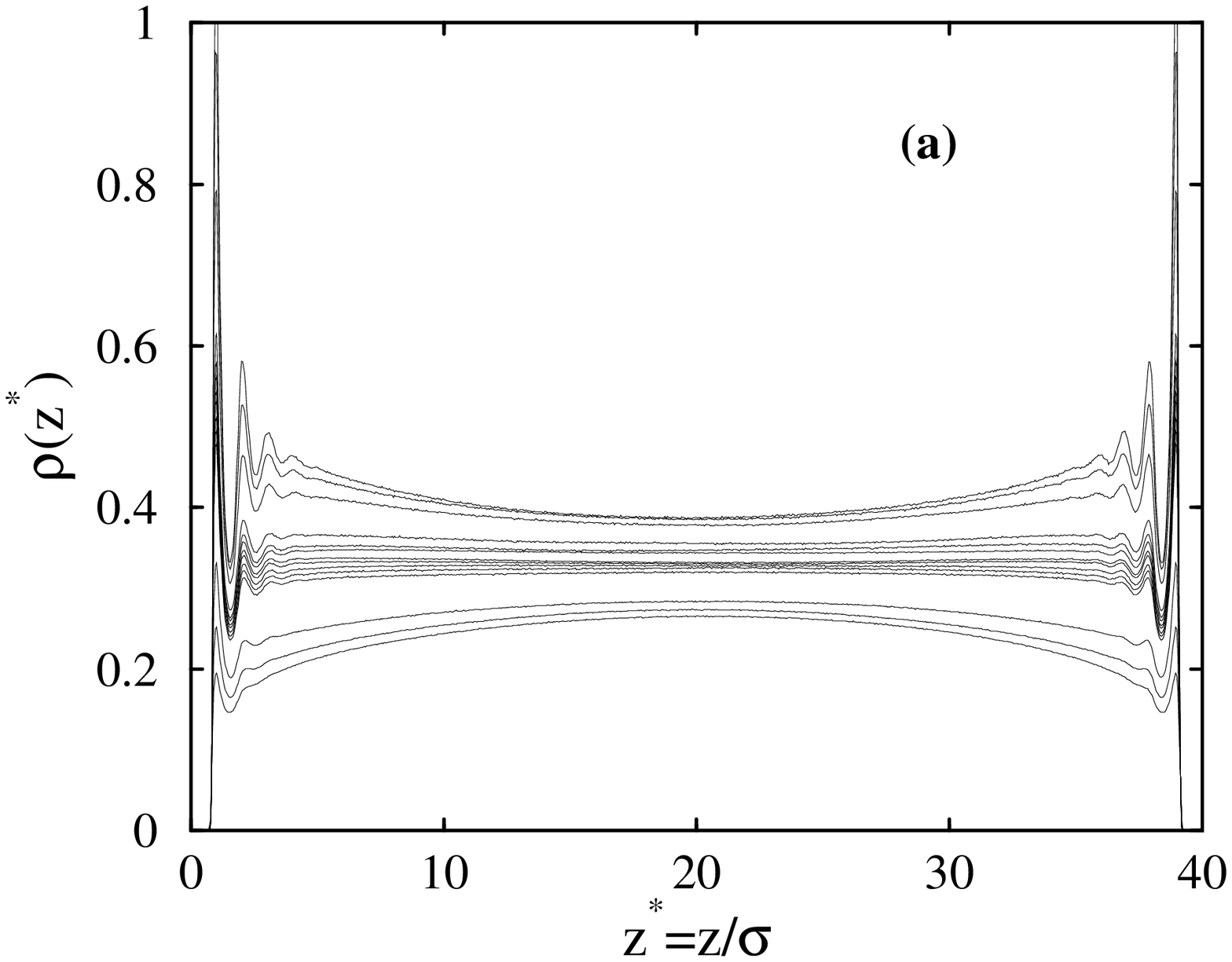}}} 
\setlength{\epsfxsize}{7.2cm} 
\centerline{\mbox{\epsffile{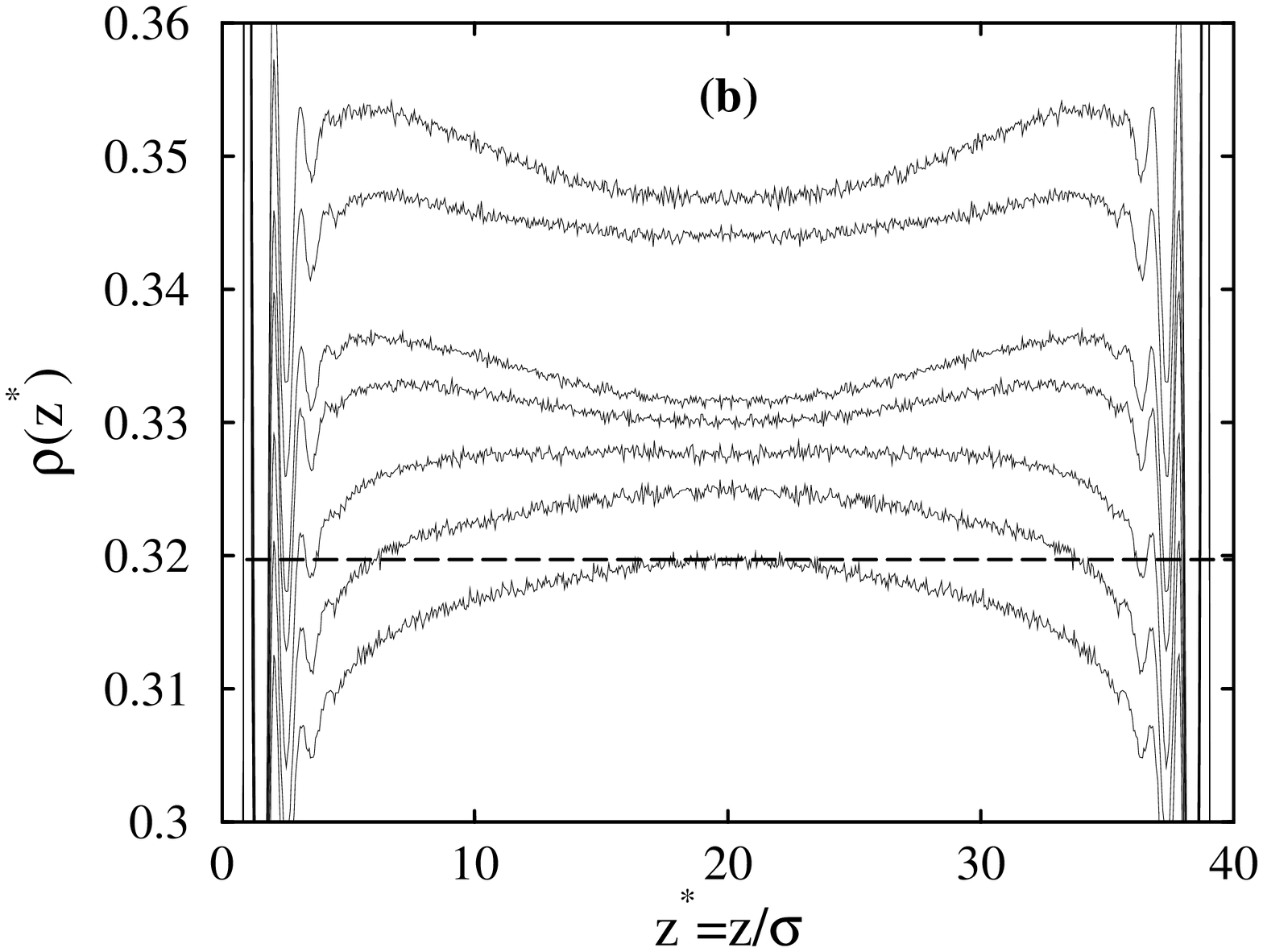}}} 
\setlength{\epsfxsize}{7.2cm} 
\centerline{\mbox{\epsffile{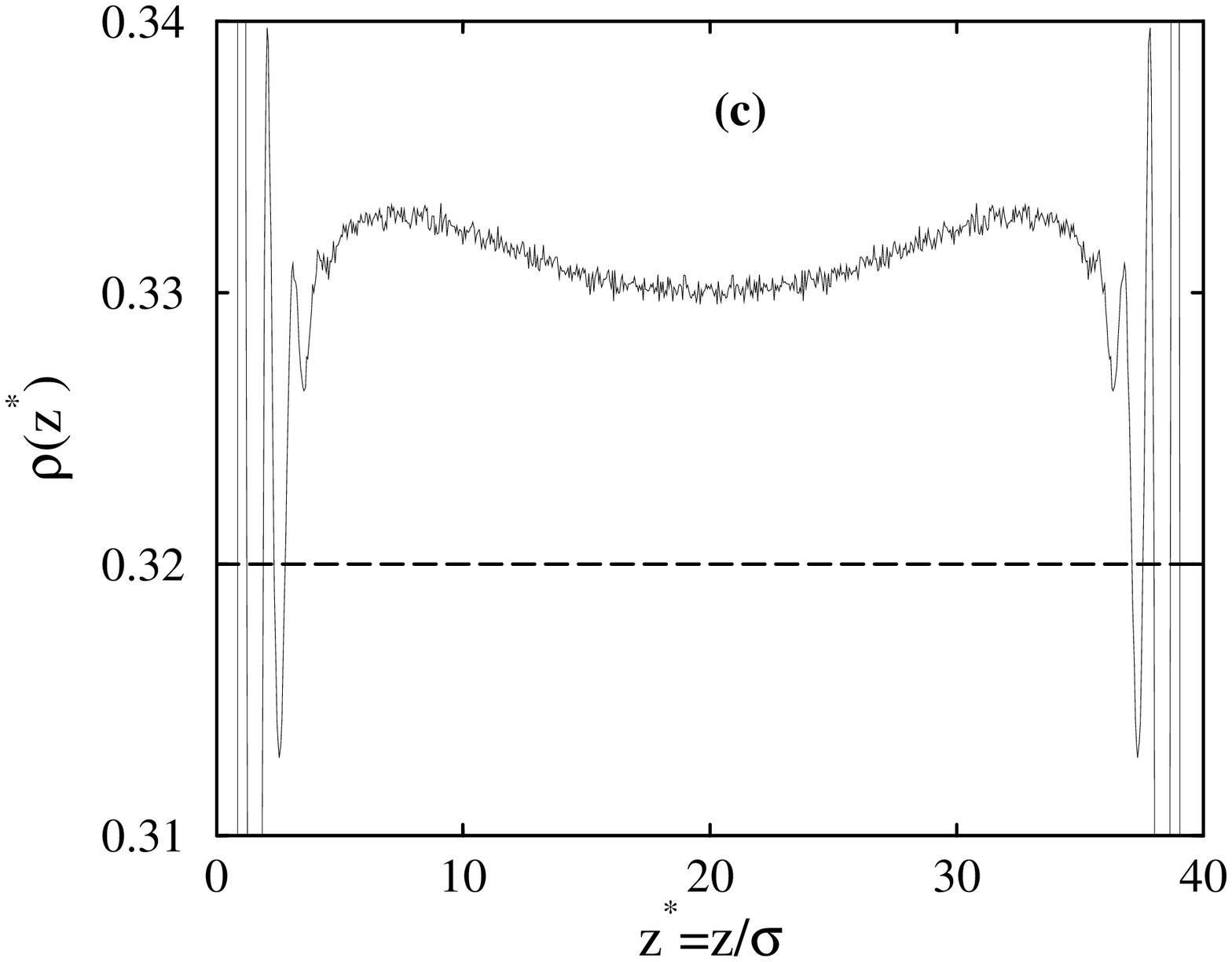}}}  
\caption{{\bf (a)} The measured critical point density profiles $\rho(z)$ 
(in units of $\sigma^3$) for 
a LJ system of size $L=15\sigma$ and width $D=40\sigma$. Wall strengths $f$ from top to bottom:
$f=$1.0, 0.9, 0.8, 0.7, 0.68, 0.67, 0.66, 0.653, 0.64, 0.63, 0.62, 0.5, 0.4, 0.3. 
The horizontal dashed line denotes the bulk critical density, $\rho_c=0.3197(4)$ 
{\bf (b)} Magnified version of (a) showing the region close to the critical 
density. Wall strengths $f$ from top to bottom:$f=$ 0.68, 0.67, 0.66, 0.653, 0.64, 0.63, 0.62.
  {\bf (c)} Magnified version of the profile for $f=0.653$ showing
 the two smooth maxima with a shallow minimum in the centre of the slit.}  
\label{fig:critprofs} 
\end{figure}

\begin{figure}[h]  
\setlength{\epsfxsize}{7.2cm} 
\centerline{\mbox{\epsffile{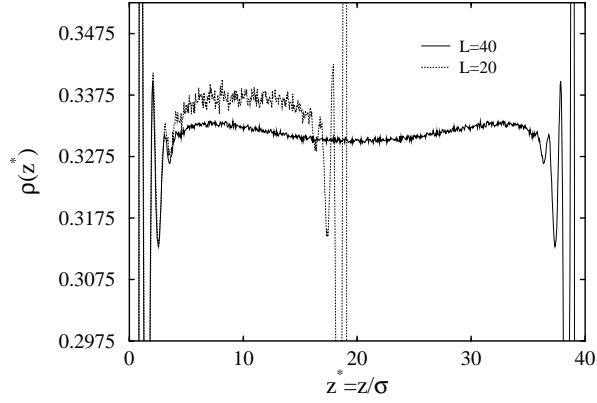}}}  
\caption{Critical point density profile $\rho(z)$ for $f=0.653$ for slits of width $D=20\sigma$ and $D=40\sigma$.
In both cases $L=15\sigma$.}  
\label{fig:fss} 
\end{figure}

\begin{figure}[h]  
\setlength{\epsfxsize}{7.2cm} 
\centerline{\mbox{\epsffile{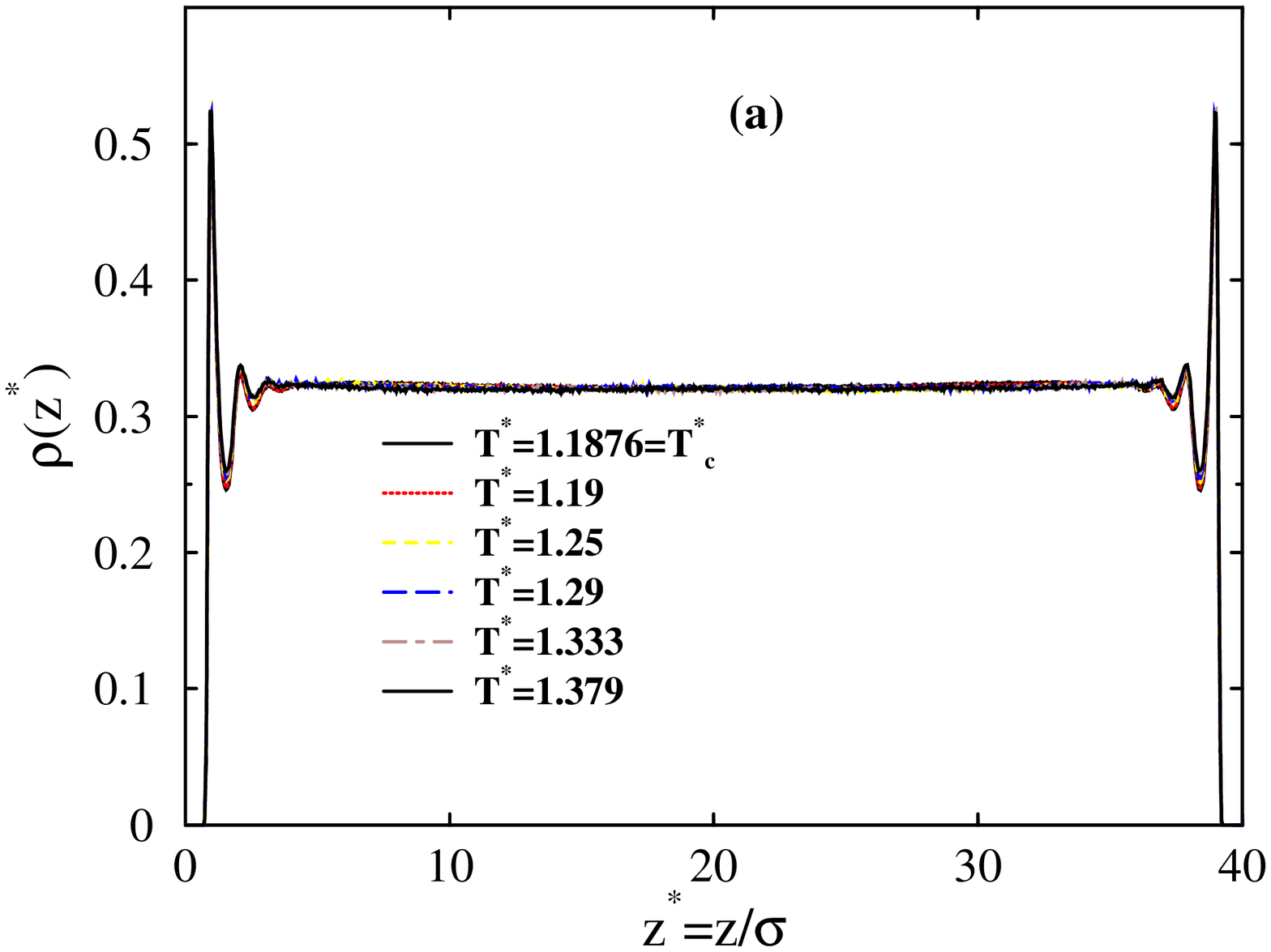}}}  
\setlength{\epsfxsize}{7.2cm} 
\centerline{\mbox{\epsffile{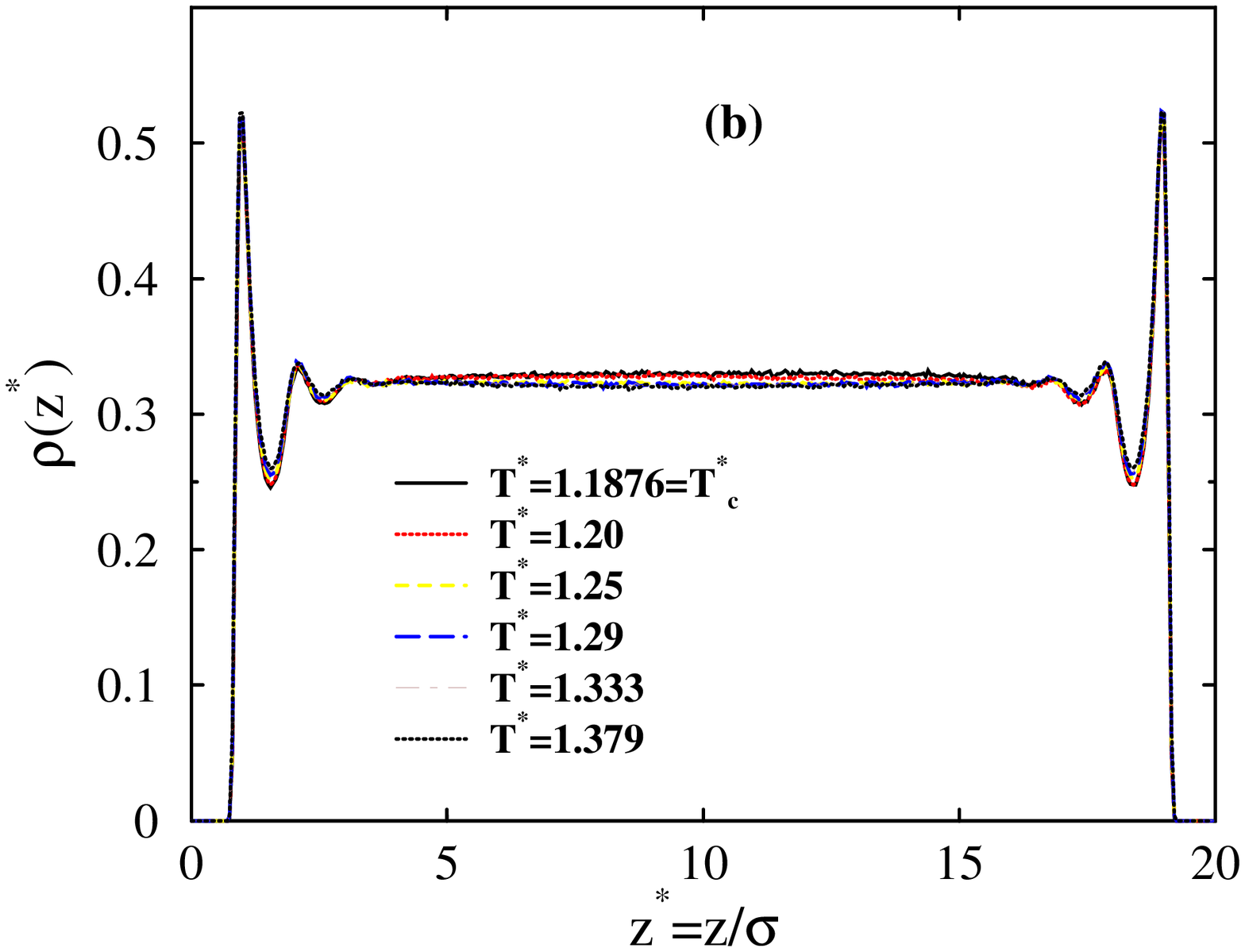}}}  
\caption{Density profiles on the critical isochore for $f=0.644$, the 'neutral' wall. (a) $L=15\sigma$  and width $D=40\sigma$
(b) $L=15\sigma$  and width $D=20\sigma$. Note how insensitive the profiles are to the value of $T^*$}  
\label{fig:neupr} 
\end{figure}

\begin{figure}[h]  
\setlength{\epsfxsize}{7.2cm} 
\centerline{\mbox{\epsffile{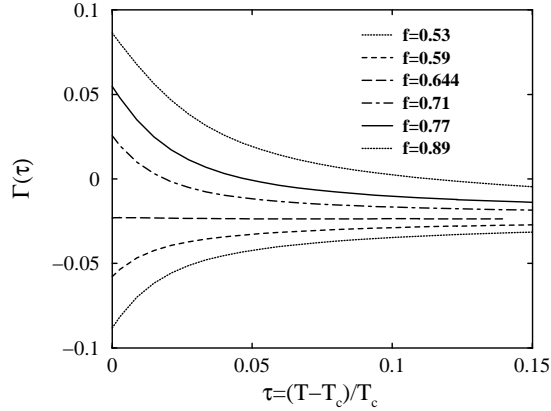}}}  
\caption{ The adsorption $\Gamma $, in units of $\sigma ^2$, calculated along
the  critical isochore as a function of reduced temperature $\tau=(T-T_c)/T_c$
for  various values of $f$. Results correspond to $L=15\sigma$
and width  $D=20\sigma$. Note that for the 'neutral' wall  $f=0.644$,
$\Gamma (\tau)$ is constant.}  
\label{fig:ads} 
\end{figure}

\begin{figure}[h]  
\setlength{\epsfxsize}{7.2cm} 
\centerline{\mbox{\epsffile{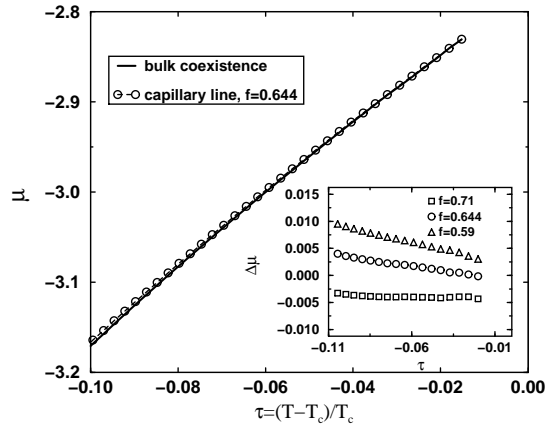}}}  
\caption{The chemical potential $\mu$ (in units of $k_BT$)
versus reduced temperature $\tau$  for capillary condensation in a system of
size $L=15\sigma$ and width $D=20\sigma$  for wall
strength $f=0.644$. The thick solid line  denotes the bulk coexistence
 line $\mu_{cx}$. The inset displays the difference 
$\Delta\mu\equiv \mu-\mu_{cx}$ between the
coexistence lines for three values of $f$. For $f=0.59$, $\mu-\mu_{cx}$
is positive, i.e. there is capillary evaporation. }
\label{fig:cap} 
\end{figure}

In this paper we have performed extensive Monte Carlo simulations
and density functional calculations of the density  profile
 and Gibbs adsorption $\Gamma$ for a Lennard-Jones fluid confined between
two planar walls under conditions where the reservoir  is at, or close to, bulk criticality. We observe that increasing the parameter $f$ that multiplies
the (10-4) wall-fluid potential alters the profile dramatically, from one typical for a purely repulsive wall  ($\Gamma$ is strongly negative) to one typical
of a strongly attractive wall ($\Gamma$ is strongly positive), 
for the critical fluid at fixed wall separation $D$.
In the simulations we find that for $f\approx 0.644$ the confined LJ
 fluid behaves  in a similar fashion to an Ising film with vanishing
 surface field $h_1=0$, i.e.
to what is observed at the ordinary transition. This 'neutral' wall scenario
is characterized by a fluid density profile which, away from the walls
 where oscillations arise, is almost constant throughout the slit and which gives rise to a $\Gamma$ that is constant  along the critical isochore 
(see Figs~\ref{fig:neupr} and ~\ref{fig:ads}).
The same value $f\approx 0.644$ gives rise to a line of capillary coexistence
for $T<T_c$ that differs very little from the bulk coexistence line
(see Fig.~\ref{fig:cap}), supporting the contention that this value corresponds
to a 'neutral' wall. Moreover, we observe that as $f$ decreases from 1
 the density profile of the critical LJ fluid, away from the walls, exhibits features reminiscent of those found for the magnetisation profile of the
critical Ising film in the crossover regime between the 'normal' and ordinary
surface universality  classes which is achieved by reducing the
 surface field $h_1$ towards zero. In particular
 for $0.7\gtrsim f \gtrsim 0.64$  and $D=40\sigma$ we find profiles
 with two smooth maxima, located away from the walls; compare
 Figs ~\ref{fig:critprofs}b and ~\ref{fig:Ispr}.
As $f$ is reduced further, or as the slit width $D$ is reduced at fixed $f$,
these maxima merge into a single maximum located at the midpoint $D/2$. 
In the Ising film  a magnetisation profile with two smooth maxima 
is characteristic of a system subject to weak surface fields $h_1=h_D$; 
the profile increases for distances 
$z\lesssim  l_1\sim h_1^{-\nu/\Delta_1^{ord}}<D/2$, before decaying to the value at the midpoint (see Sec.~\ref{sec:Is}). For weaker $h_1$ (or smaller $D$)
a single maximum forms at the midpoint when $l_1\sim D/2$.

In the  light of the simulation results it is tempting to argue that the near critical LJ fluid does exhibit several features in common with the Ising system.
Consideration of the adsorption and of capillary condensation would suggest
identifying $h_1=f-f_{neutral}$, with $f_{neutral}\approx 0.644$.
However, the fluid density profiles at the critical point clearly do not exhibit the symmetry 
$m(z;-h_1)=-m(z;h_1)$ of the magnetisation profile of the Ising system - even 
when some attempt is made at coarse graining the oscillations near the walls
and subtracting  a constant density. Thus it is not obvious what is
 the best choice of order parameter profile for the fluid.
 The form of the confining potential obviously plays an important role
and in order to investigate this further we performed a series of 
DFT calculations using a very simple WDA to
 incorporate short-ranged correlations.

\begin{figure}[h]  
\setlength{\epsfxsize}{7.2cm} 
\centerline{\mbox{\epsffile{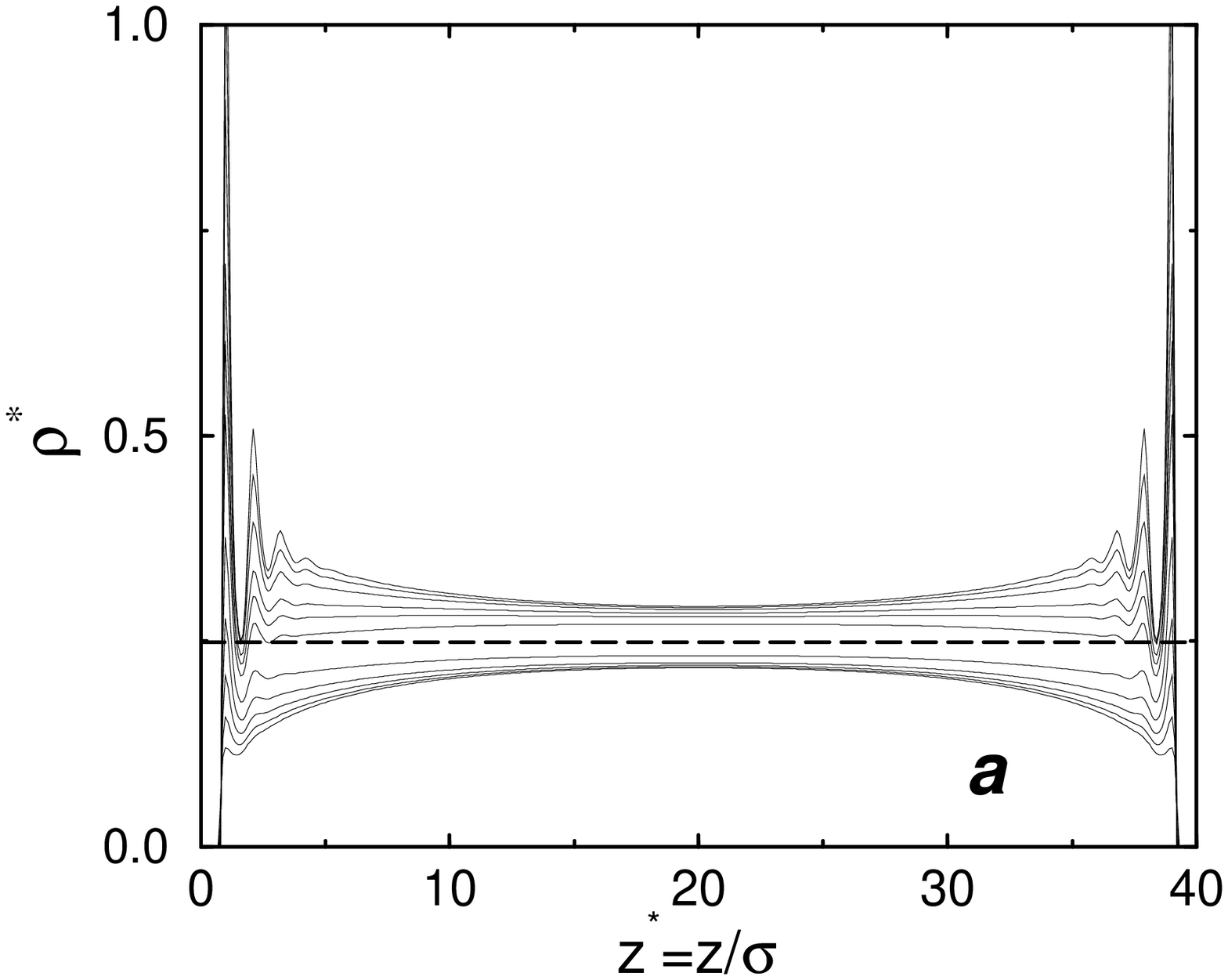}}}  
\setlength{\epsfxsize}{7.2cm} 
\centerline{\mbox{\epsffile{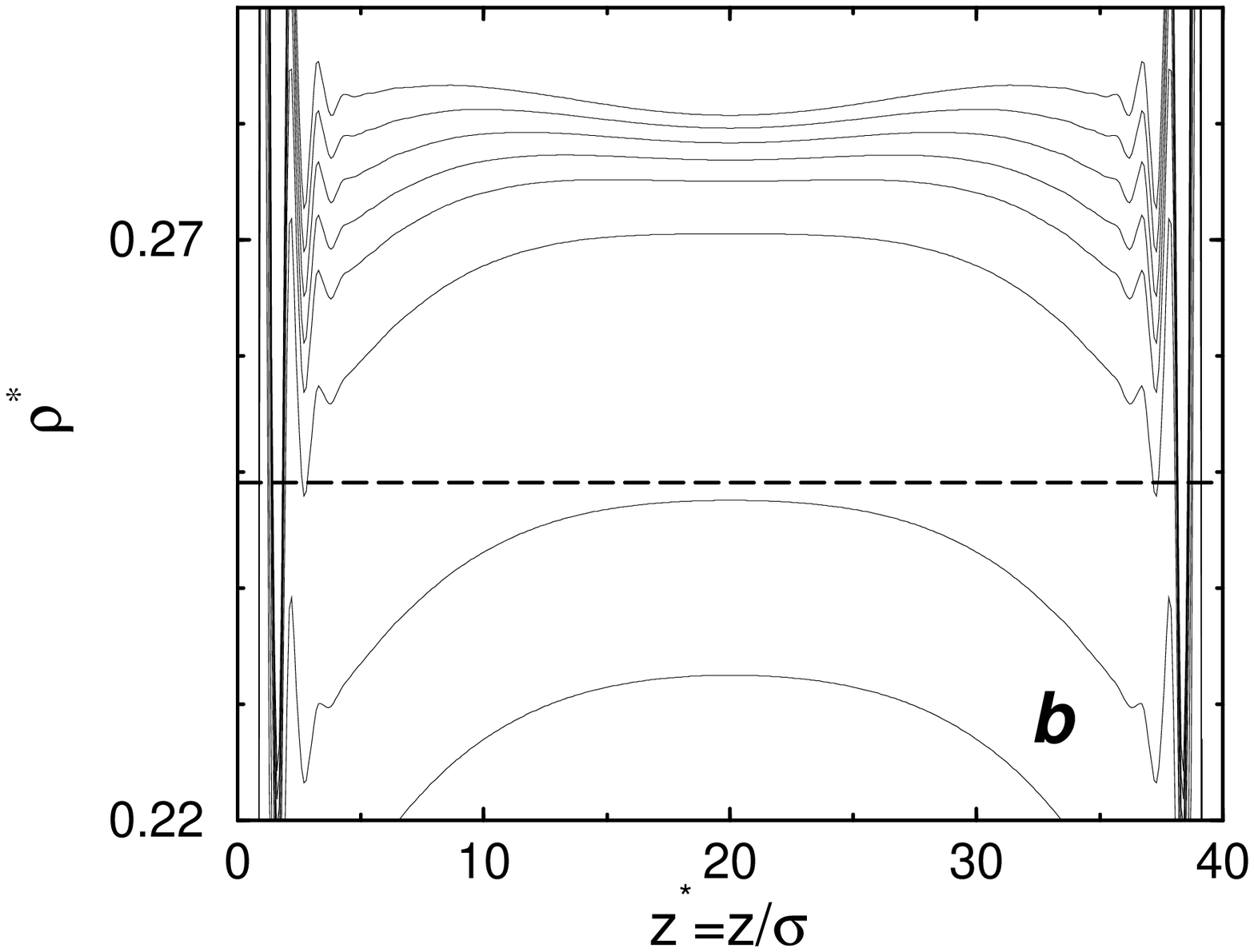}}}  
\caption{Critical point density profiles $\rho (z)$ (in units of $\sigma ^3$ )
for a slit of width $D=40\sigma$
and  the long-ranged wall-fluid potential $U_w(z)$ (\ref{eq:extlr}) 
 obtained using the WDA
 for several
values of the wall strength  $f$:
{\bf (a)} from  top to bottom:
$f=$1.0,  0.9, 0.8, 0.7, 0.65, 0.6, 0.5, 0.4, 0.3, 0.2, 0.1.
{\bf (b)} Magnified version for $f$= 0.66, 0.65, 0.64, 0.63, 0.62, 0.6, 0.55
 and 0.5. Note that for $0.62\lesssim f\lesssim 0.68$ two smooth
maxima, separated by a shallow minimum at the midpoint, arise at distances
away from the walls, outside the region  where the oscillations due
 to packing occur.
The horizontal dashed line denotes the bulk critical density,
 $\rho _c=0.24912(9)$. }  
\label{fig:WDALR} 
\end{figure}

\begin{figure}[h]  
\setlength{\epsfxsize}{7.2cm} 
\centerline{\mbox{\epsffile{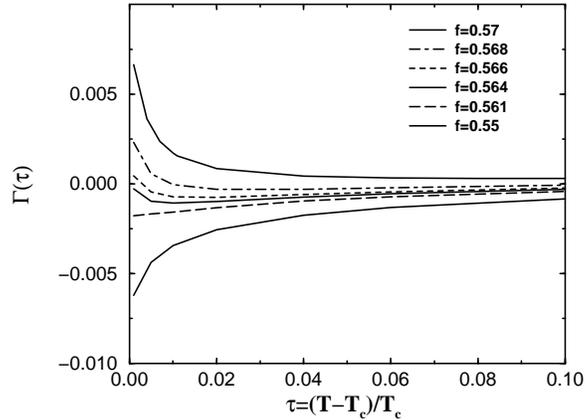}}}  
\caption{WDA results for the adsorption $\Gamma$, in units of $\sigma ^2$,
calculated for a slit of width $D=40\sigma$ along the critical
 isochore as a function of reduced temperature $\tau$, for several
 values of $f$. The wall-fluid
potential $U_w(z)$ (\ref{eq:extlr}) is long-ranged. Note that the adsorption
 is very small and only weakly dependent on temperature for the range
 of values of $f$ shown here. }
\label{fig:ads1} 
\end{figure}

\begin{figure}[h]  
\setlength{\epsfxsize}{7.2cm} 
\centerline{\mbox{\epsffile{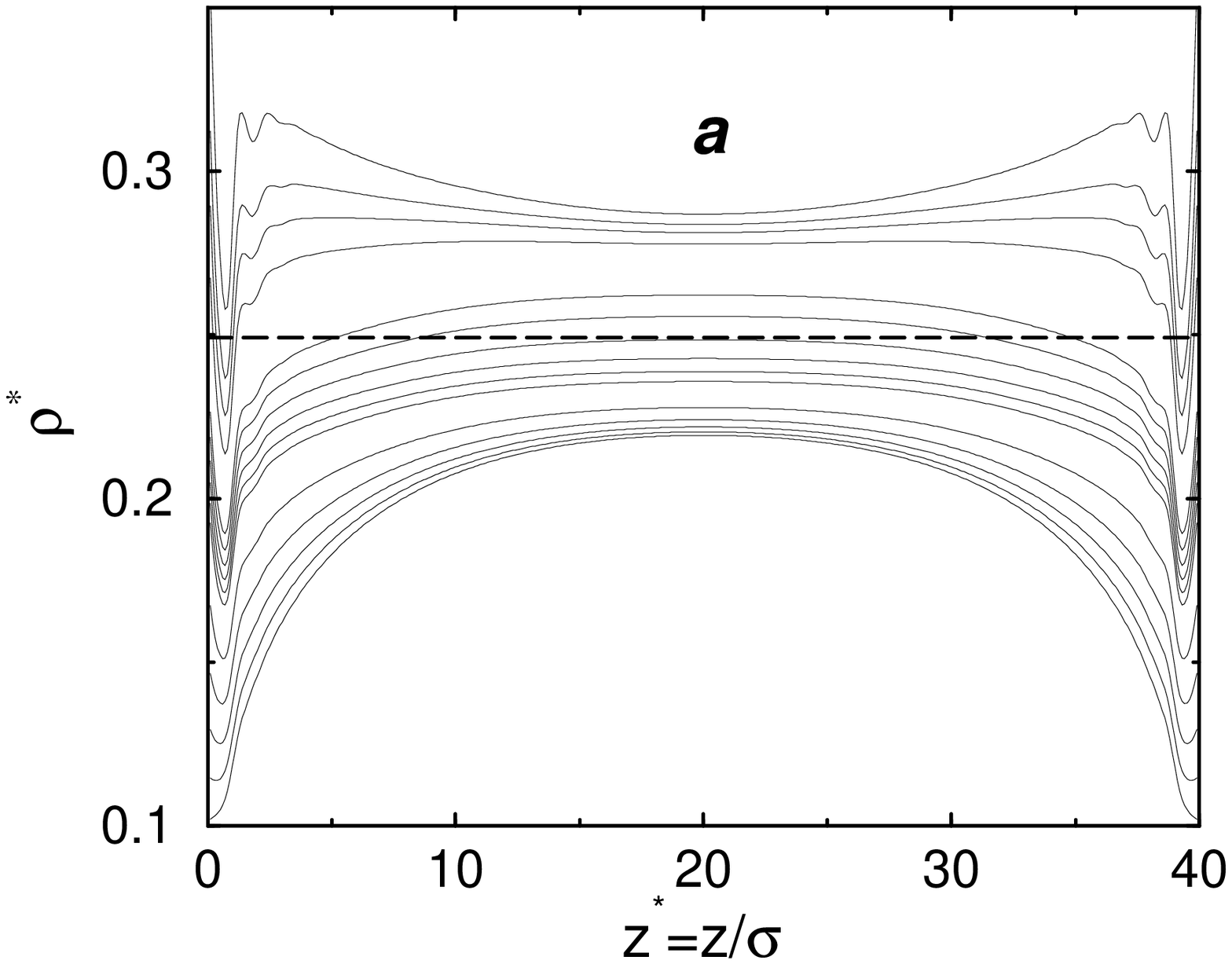}}}  
\setlength{\epsfxsize}{7.2cm} 
\centerline{\mbox{\epsffile{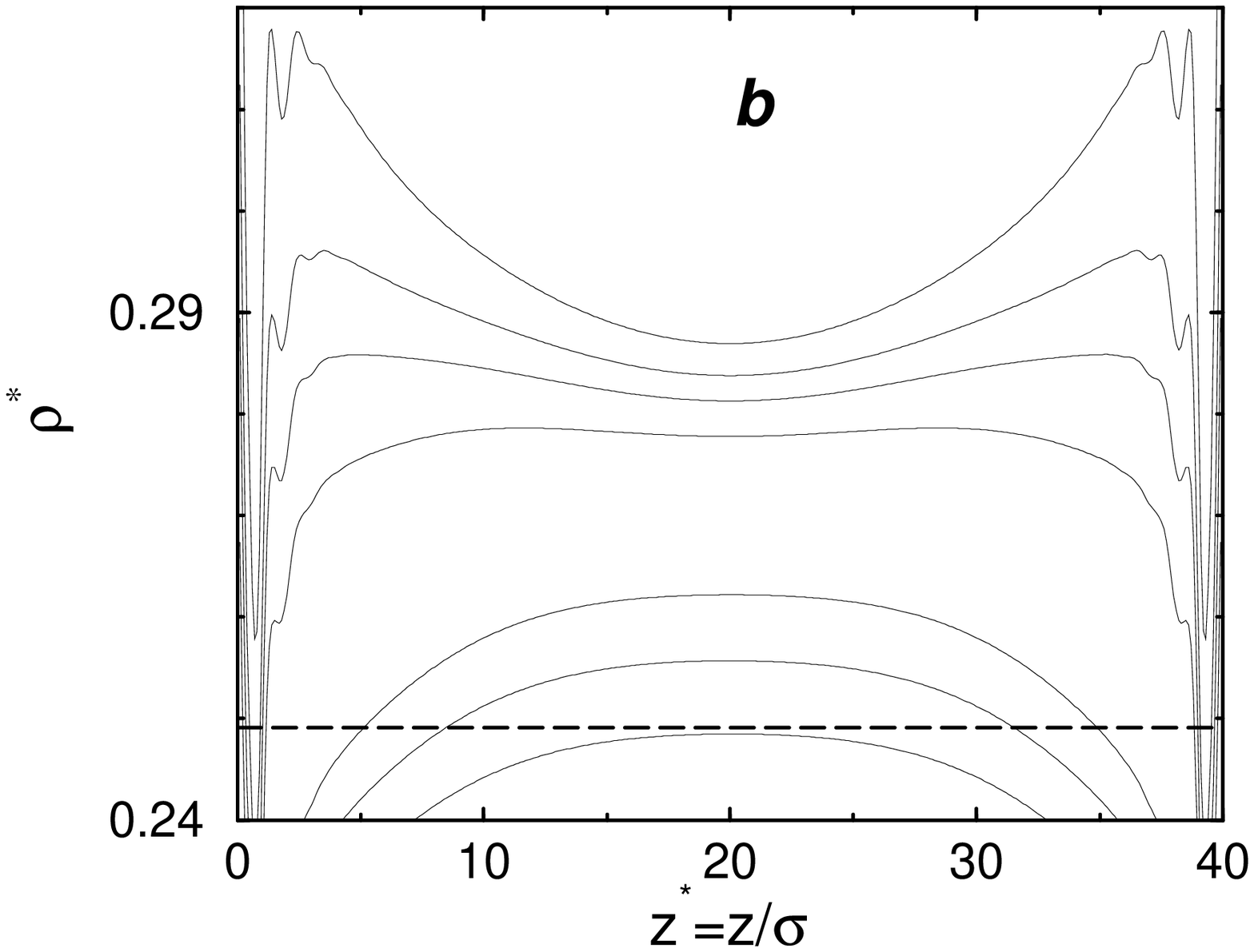}}}  
\caption{Critical point density profiles
 $\rho (z)$ (in units of $\sigma ^3$ )
for a slit of width $D=40\sigma$
and  the short-ranged wall-fluid potential $U_{ws}(z)$ (\ref{eq:extsr})
 obtained using the  WDA for several values of
 the wall strength $f$: {\bf  (a)} from  top to bottom:
$f=$1.0,  0.9, 0.85, 0.8, 0.7, 0.68, 0.66, 0.64, 0.62, 0.6, 0.5, 0.4, 0.3,
 0.2, 0.1.
{\bf (b)} Magnified version  of {\bf (a)}  for the largest values of $f$
: 1, 0.9, 0.85, 0.8, 0.7, 0.68, 0.66.
The horizontal dashed  line denotes the bulk critical density, $\rho _c=0.24912(9)$. 
}  
\label{fig:WDASR} 
\end{figure}

\begin{figure}[h]  
\setlength{\epsfxsize}{7.2cm} 
\centerline{\mbox{\epsffile{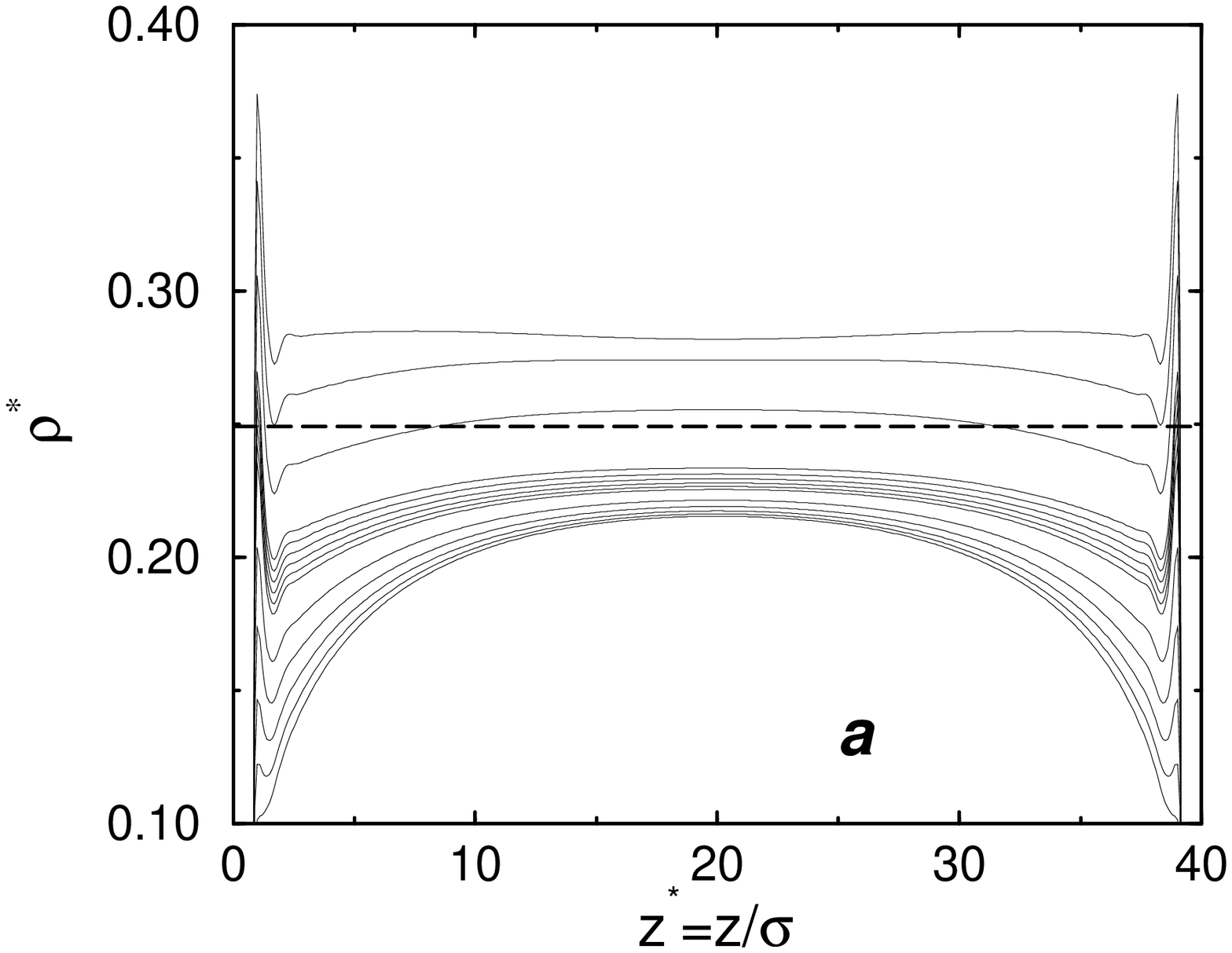}}}    
\setlength{\epsfxsize}{7.2cm} 
\centerline{\mbox{\epsffile{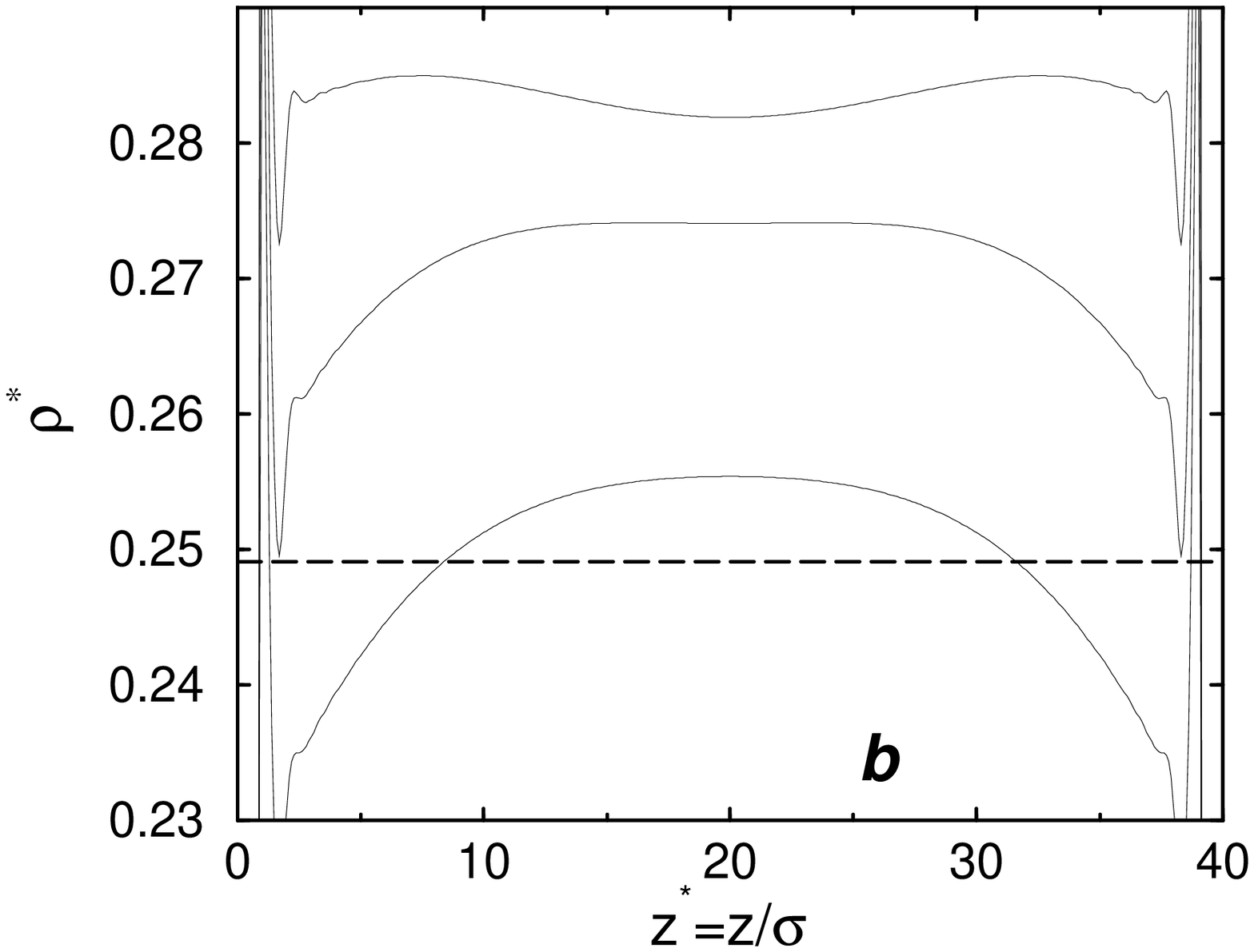}}}  
\caption{Critical point density profiles
 $\rho (z)$ (in units of $\sigma ^3$ )
for a slit of width $D=40\sigma$
and  the long-ranged wall-fluid potential $U_w(z)$  (\ref{eq:extlr})  obtained using the  LDA
 for the same choice of 
  wall strengths $f$ as in Fig.\ref{fig:WDASR}(a).
{\bf (b)} Magnified version  for $f=1, 0.9, 0.8$.
The horizontal dashed  line denotes the bulk critical 
density, $\rho _c=0.24912(9)$. 
}  
\label{fig:LDA} 
\end{figure}

Our DFT results also point to the existence of a 'neutral' wall;
 the adsorption $\Gamma (\tau)$ calculated along the critical isochore 
is almost flat for $f$ lying between 0.561 and 0.564 pertaining to the same 
(10-4) wall potential and wall separation $D=40\sigma$ as the simulations 
- see Fig.~\ref{fig:ads1}. Moreover, the DFT density profiles exhibit
the characteristic two smooth maxima for an intermediate range of $f$ observed in the simulations - see Fig.~\ref{fig:WDALR}.
What does this lead us to conclude about the role of critical fluctuations in determining the shape of the density profiles for 'weak' surface fields?
Recall that unlike simulations DFT is a mean-field treatment which does not
incorporate effects of fluctuations. 
The critical scaling arguments of Sec.~\ref{sec:Is} state categorically that
the mean field order parameter profile for the Ising system subject 
to a surface field $h_1$ acting on the surface layer only 
should vary monotonically away from the wall, 
i.e. there should be no maxima. The mean-field value of the exponent 
that describes the short-distance behaviour of the profile,
 Eq.~(\ref{eq:mag}), is $\kappa=0$. That we find smooth maxima 
in the density profiles  within our DFT treatment  would seem to 
reflect  the shape of the confining (wall-fluid)
potential. Note that 
the precise form of wall-fluid potential is not crucial as we also find
the maxima, albeit for larger values of $f$, for a short-ranged, exponentially
decaying attractive potential - see Fig.~\ref{fig:WDASR}. There
are quantitative differences between DFT and simulation.
 The maxima are more pronounced, are located closer to the walls and appear
to shift faster towards the midpoint as $f$ is decreased in simulation 
and this might reflect the presence of fluctuation effects.
The latter give rise to a slightly higher value of the exponent determining 
the length scale $l_1\sim h_1^{-\nu/\Delta _1^{ord}}$ than in mean field, i.e. the $d=3$ Ising value is 1.3 rather than 1. These observations suggest that one should re-analyse the arguments which lead to Eq.~(\ref{eq:mag}). Both arguments given in Sec.~\ref{sec:Is}A  are based on the assumption that the surface magnetisation  $m_1$ and $m(z\sim 0)$ respond linearly to a weak $h_1$ for $T=T_c$. Whilst this is eminently  plausible for the Ising system it is not clear
how to translate to the case of a fluid near a wall. It is possible that the shape of the external potential has non-trivial repercussions for the short-distance expansion of the order parameter profile. However, one should also bear in mind that 
there is a further complication in attempting to relate crossover
behaviour in fluids to that in systems with Ising symmetry. The reduced
order parameter symmetry of bulk fluids lead to scaling field mixing
\cite{REHR,KIM}. However, to our knowledge, there have to date been no
considerations of the consequences of this reduced symmetry in the
context of surface critical behaviour of fluids. Any scaling theory
that seeks to encompass fluid and magnet surface critical behaviour
within a unified description must address this issue. 

\begin{figure}[h] 
\setlength{\epsfxsize}{7.2cm}
\centerline{\mbox{\epsffile{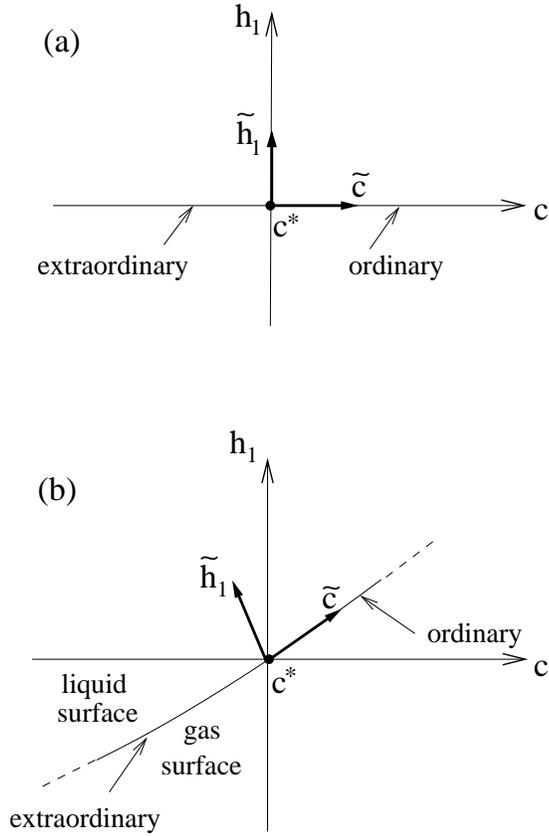}}}

\caption{The form of the surface scaling fields $\tilde c$ and  $\tilde
h_1$, for a system at its bulk critical point parameters, as described in
the text. {\bf (a)} A model of the Ising symmetry where there is no
mixing, so $\tilde{c}=c$ and $\tilde h_1=h_1${\bf (b)} A fluid
system. }
\label{fig:fields}
\end{figure}

One way of doing so may be to generalize the role of the surface
scaling fields $c$ and $h_1$ (defined for Ising-like systems) such that
regardless of the model system, they correspond to equivalent paths in
the surface phase diagram. Consider the case of an Ising model in zero
bulk field at its bulk critical temperature. Within the surface phase
diagram, the line $h_1=0$ corresponds to a special symmetry line of the
system because the magnitude of the surface order parameter is
invariant with respect to a change of sign of $h_1$. The scaling field
$c$ (which relates to missing neighbours and surface induced
modifications of surface couplings), controls the temperature at which
the surface orders relative to the bulk. The sign or magnitude of $c$
does not determine which phase is preferred at the wall, thus the $c$
field  can be
represented as a vector coincident with the symmetry line $h_1=0$
(Fig.~\ref{fig:fields}(a)).

By contrast in a fluid, missing neighbours typically lead to a
segregation of one phase to the wall, i.e. to breaking of the order
parameter symmetry with respect to its bulk critical point value. In
terms of the geometrical picture outlined above, the line of surface
order parameter symmetry lies off-axis in the surface phase diagram and
does not coincide with the c-field. In attempting to restore the
symmetry, one might consider applying a non-zero external surface field
$h_1$ that couples to the order parameter near the surface, and whose
magnitude is chosen such as to return the value of the order parameter
at the surface to its bulk critical value.  Within this framework, the
combined $c$ and $h_1$ fields can be regarded as an {\em effective}
field $\tilde{c}$ directed along the symmetry line. Similarly we
require an effective field  $\tilde{h}_1$ (the scaling equivalent to
the Ising model $h_1$) which measures deviations from this line
(Fig.~\ref{fig:fields}(b)). Formally these ends can be attained by
introducing `mixed' surface fields having the form

\begin{eqnarray}
{\tilde c} &=& (c-c^\star)+a_1h_1\\\nonumber
{\tilde h}_1 &=& h_1+a_2 (c-c^\star)\;,
\label{eq:mixedfields}
\end{eqnarray}
with $a_1$ and $a_2$ system specific constants (both zero in
the Ising model context) whose values control the degree of field mixing.

The starting point for exploring the consequences of field mixing for
surface critical behaviour is an expression for the singular part of the surface
free energy. In terms of the mixed fields  this is

\begin{equation}
F^s_{sur}\approx \tilde\tau^{2-\alpha}f^\pm_{sur}(\tilde{\mu}\tilde{\tau}^{-\Delta},\tilde{c}\tilde{\tau}^{-\phi},\tilde{h}_1\tilde{\tau}^{-\Delta_1})\;,
\label{eq:mixedfe}
\end{equation}
where $f^\pm_{sur}$ denote scaling functions, while $\tilde{\tau}$ and
$\tilde{\mu}$ are (mixed) bulk scaling fields which in the Ising
context reduce to the reduced temperature and reduced bulk field
respectively \cite{KIM}. The positive constants $\Delta$ and $\Delta_1$
are respectively the gap exponent and surface gap exponents, while
$\phi$ is an exponent controlling the crossover from the ordinary to
the normal universality classes.

We shall not attempt to explore the consequences of surface field
mixing in any great depth here, although we do hope to report on the
matter in future work. For the time being, we merely point out that one
corollary of Eq.(\ref{eq:mixedfe}) is that--in a simple fluid--the
scaling equivalent of the Ising model magnetisation profile $m(z)$ is
not the density  profile $\rho(z)$, but a profile formed by linearly
combining the density and the energy density profiles. One should also
consider the adsorption not on the critical isochore, but on an
analytical extension of the coexistence diameter. These observations
may prove important in exposing more fully the universality of surface
critical phenomena for fluids and magnets.

Finally, with regard to the applicability of the above considerations
to realistic fluids, the question arises as to just what constitutes
the surface field $h_1$ in a fluid. Clearly the wall-fluid potential in
general couples to the order parameter in the vicinity of the surface
and thus can be used to tune its value. However, as noted in
Sec.~\ref{sec:intr}, a realistic potential such as the 10-4 potential
used in this work will not allow one to restore the order parameter to
its bulk critical value exactly at all distances from the wall. Short
ranged packing effects will always be visible close to the wall.
Nevertheless, in the context of the ordinary transition, the length
scale on which packing effects occurs will be negligible compared to
the length scale of the critical phenomena (set by $l_1$). Hence one
might speculate that fluid-magnet universality should be evident
provided one views the system on sufficiently large length scales, i.e.
such that the value of a suitably {\em coarse-grained} order parameter
profile matches its bulk value near the wall.

\acknowledgements 
We have benefited from conversations with A. Ciach, S. Dietrich, M. Krech, B. M. Law, A. O. Parry, F. Schlesener and, in particular, P. J. Upton.
R. E. is grateful to S. Dietrich and his colleagues at MPI Stuttgart for their hospitality and to the Alexander von Humboldt Foundation for financial support
under GRO/1072637. This work was supported by Royal Society travel grants
and partially by KBN (grant no. 4 T09A 066 22).
 
\begin{thebibliography}{99}
\bibitem{footnote} The surface layer magnetisation $m_1\sim |\tau|^{\beta _1^{ord}}$ as $\tau \equiv (T-T_c)/T_c\to 0^{-}$,
with $\beta _1^{ord}\sim 0.8$ for the Ising model in $d=3$. The exponents $\beta _1^{ord}$ and $\Delta _1^{ord}$
are related via $\beta _1^{ord}=(d-1)\nu-\Delta _1^{ord}$.
\end{thebibliography}
\end{document}